# Electron-Phonon Interaction in Tetrahedral Semiconductors

**Manuel Cardona**[*]

*Max-Planck-Institut für Festkörperforschung, Heisenbergstrasse 1,
70569 Stuttgart, Germany*

**Abstract**

Considerable progress has been made in recent years in the field of *ab initio* calculations of electronic band structures of semiconductors and insulators. The one-electron states (and the concomitant two-particle excitations) have been obtained without adjustable parameters, with a high degree of reliability. Also, more recently, the electron-hole excitation frequencies responsible for optical spectra have been calculated. These calculations, however, are performed with the constituent atoms fixed in their crystallographic positions and thus neglect the effects of the lattice vibrations (i.e. electron-phonon interaction) which can be rather large, even larger than the error bars assumed for *ab initio* calculations.

Effects of electron-phonon interactions on the band structure can be experimentally investigated in detail by measuring the temperature dependence of energy gaps or critical points (van Hove singularities) of the optical excitation spectra. These studies have been complemented in recent years by observing the dependence of such spectra on isotopic mass whenever different stable isotopes of a given atom are available at affordable prices. In crystals composed of different atoms, the effect of the vibration of each separate atom can thus be investigated by isotopic substitution. Because of the zero-point vibrations, such effects are present even at zero temperature ($T = 0$).

In this paper we discuss state-of-the-art calculations of the dielectric function spectra and compare them with experimental results, with emphasis on the differences introduced by the electron-phonon interaction. The temperature dependence of various optical parameters will be described by means of one or two (in a few cases three) Einstein oscillators, except at the lowest temperatures where the $T^4$ law (contrary to the Varshni $T^2$ result) will be shown to apply. Increasing an isotopic mass increases the energy gaps, except in the case of monovalent Cu (e.g., CuCl) and possibly Ag (e.g., AgGaS$_2$). It will be shown that the gaps of tetrahedral materials containing an element of the first row of the periodic table (C,N,O) are strongly affected by the electron-phonon interaction. It will be conjectured that this effect is related to the superconductivity recently observed in heavily boron-doped carbon.

---

[*] m.cardona@fkf.mpg.de, Fax: +49-711-689-1712

# 1. Introduction

Calculations of the band structure of tetrahedral semiconductors first appeared in the mid 1950's [1]. They were highly instrumental in the understanding of experimental results which appeared at a fast rate because of the technological importance of these materials. Foremost among these results we mention cyclotron resonance, discovered 50 years ago [2-4]. The early calculations, often using the local density approximation, agreed only semi-quantitatively with the experimental results. This fact led to the development of semiempirical techniques in which several parameters were adjusted so as to fit rather accurately a number of experimental results. The most popular among semiempirical techniques is based on the concept of pseudopotentials [5]. In the mid 1970's the semiempirical pseudopotentials were replaced by *ab initio*, self-consistent versions [5]. They enabled the calculation of structural and mechanical properties, such as the lattice parameters, the elastic constants, phase transitions under pressure and even the phonon dispersion relations [6]. Nevertheless, they failed to correctly reproduce the excitation energies involved in the optical response spectra [7] which were well reproduced by the semiempirical techniques (using adjustable parameters, of course). Agreement between calculated and measured energy gaps was restored by explicitly including correlation corrections to the electronic states involved, by means of the so-called GW techniques [8]. Thus far, however, no successful attempts were made to reproduce *ab initio* the experimental excitation spectra which were represented reasonably well by the earlier semiempirical calculations.

*Ab initio* calculations of the linear optical response, i.e., of the complex dielectric function $\varepsilon(\omega)$, must include not only correct single particle states but also the electron-hole interaction in the excited state, usually referred to as the exciton interaction.

*Ab initio* calculations of $\varepsilon_2(\omega)$ [the imaginary part of $\varepsilon(\omega)$] were performed by a number of techniques in the late 1990's [9]. Although these calculations agreed semi-quantitatively with measurements performed at low temperature [10], they implicitly assumed that the atoms are placed at their fixed crystallographic positions, disregarding the quantum-mechanical fact that they vibrate even at $T = 0$ K (the so-called zero-point vibrations). These vibrations, i.e., the phonons, can strongly affect the one-particle and the two-particle electronic excitation spectra even at $T = 0$ K and even more so at higher temperatures [11].

The strong temperature dependence of $\varepsilon_2(\omega)$ is exemplified in Fig. 1 for germanium [12]. Note that at around 1.5 eV $\varepsilon_2$ increases with temperature by an order of magnitude. Somewhat smaller, but still large changes in $\varepsilon_2(\omega)$ with temperature are observed throughout most of the spectra. Related changes with isotopic mass have been observed recently [13].

Such effects will be the main subject of this "Research Article". I dedicate it to Elias Burstein in honor of his contributions to the field, both as a creative researcher [3] and as a resourceful mentor, and also as a friend. He founded Solid State Communications 41 years ago. He led the journal with a steady hand till I took over in 1992. After his retirement, he was always available to me with advice and encouragement. On the occasion of my retirement as Editor-in-Chief (to take place in Dec. of 2004), I would also like to wish Aron Pinczuk, the new Editor-in-Chief, a great success in this challenging but not always easy endeavor.

## 2. Effect of electron-phonon interaction on electronic states and electron-hole excitations

As already mentioned, no calculations reproducing temperature-dependent spectra, such as those of Fig. 1, have thus far been performed. Extant calculations, either of the semiempirical [5] or the *ab initio* [9,10] variety, should be compared with measurements at low temperature $T \ll T_D$, (where $T_D$ is the Debye temperature) but even then, the effect of the zero-point vibrations is being implicitly neglected. This effect is, however, not negligible as can be easily surmised from Fig. 1. For Ge, with $T_D \simeq 350$ K, the difference between low temperature (~100 K) and the lattice at rest should be similar to the difference between 100 K and $T_D$, i.e., a factor of two in $\varepsilon_2(\omega)$ for $\omega \simeq 2$ eV.

Although no calculations of $\varepsilon_2(\omega,T)$ are available, considerable work has appeared concerning the temperature dependence of the critical point structure displayed in these spectra ($E_1$, $E_1 + \Delta_1$, $E'_0$, $E_2$ and $E'_1$, plus $E_0$ and $E_{ind}$, which have not been shown in Fig. 1) [16,17]. The present article is concerned with the temperature dependent shifts induced in the critical point energies by the electron-phonon interaction and their broadenings, which correspond to the imaginary part of the interaction self-energy [18].

One usually distinguishes three contributions to the electron-phonon interaction effects. The simplest one is indirect and related to the thermal expansion [19]. It is based on the so-called quasi-harmonic approximation which assumes that the interatomic potential is harmonic, but the lattice constant depends on temperature (including zero-point effects [20]). The corresponding effect on a given critical point energy (i.e., an energy gap) is then written as:

$$\Delta\omega_g(T) = a \cdot \left(\frac{\Delta V(T)}{V_0}\right)_p = \left(\frac{\partial \omega_g}{\partial p}\right)_T \cdot B_0 \cdot \left(\frac{\Delta V(T)}{V_0}\right)_p \tag{1}$$

where a is the volume deformation potential, $p$ the pressure, $B_0$ the bulk modulus and $\Delta V(T)$ the volume expansion of the quasiharmonic approximation, including zero-point effects. Note that $B_0$ as well as $\left(\partial \omega_g/\partial p\right)_T$ depend only weakly on temperature. The strongest temperature variation in Eq. (1) appears through the thermal expansion $\left(\Delta V(T)/V_0\right)_p$. This term will be discussed in Sect. 3.

Beside the quasiharmonic effect of Eq. (1), two additional terms, representing explicitly the electron-phonon interaction, contribute to the temperature dependence of electronic states. They are given by the Feynman diagrams of Fig. 2. These terms are of second order in the phonon amplitude $u$. Diagram (a) is called a self-energy diagram and each of the two dots represents the first-order electron-phonon interaction which is taken twice, in second-order perturbation theory (hence, the $u^2$ dependence). Diagram (b) represents the interaction of an electron with two phonons, i.e., the electron – two-phonon interaction. Higher order terms in $u$ are not needed to interpret most of the extant experimental results concerning the dependence of electronic excitations on $T$, including zero-point effects.

The diagram of Fig. 2b is the simplest one to evaluate, as first suggested by Antončík [21]. In the local pseudopotential formulation it corresponds to "screening" the various components of the ionic pseudopotential $v(\mathbf{G})$, where $\mathbf{G}$ are the reciprocal lattice vectors, by Debye-Waller factors of the form:

$$e^{-\frac{1}{2}\langle u^2\rangle|\mathbf{G}|^2}=1-\frac{1}{2}\langle u^2\rangle|\mathbf{G}|^2+\cdots \tag{2}$$

In this manner, the temperature effect appears in the thermodynamical expectation value $\langle u^2\rangle$ which is given by (we consider for simplicity a cubic crystal with two equal atoms per primitive cell. $u$ is the displacement of an atom along a given coordinate axis)

$$\langle u^2\rangle=\frac{\hbar}{4M\Omega}\langle 1+2n_B\rangle, \tag{3}$$

where $M$ is the atomic mass (or the average isotopic mass if several isotopes are present) and $n_B$ is the Bose-Einstein factor of each individual phonon:

$$n_B(\Omega,T)=\frac{1}{e^{\frac{\Omega}{T}}-1} \tag{4}$$

where $\Omega$ is the phonon frequency in the same units as the temperature $T$. In the high temperature limit we find

$$\langle u^2\rangle=\frac{\hbar T}{2M\langle\Omega\rangle^2} \qquad (T \text{ and } \Omega \text{ in units of frequency}) \tag{5}$$

where $\langle\Omega\rangle$ is an average frequency which should be close to the Debye temperature $T_D$ and is proportional to $M^{-\frac{1}{2}}$. According to Eq. (5) $\langle u^2\rangle$ in the high temperature limit <u>is independent of $M$</u>. At low temperatures ($T\ll T_D$) we find from Eq. (3) the zero-point vibration amplitude:

$$\langle u^2\rangle=\frac{\hbar}{4M\langle\Omega\rangle}\propto M^{-\frac{1}{2}} \tag{6}$$

The zero-point amplitude ($\propto M^{-\frac{1}{2}}$) thus decreases with increasing $M$ and becomes negligible for $M\to\infty$, as expected for a quantum effect. Correspondingly, for $T\gg T_D$ the dependence on $M$ disappears.

The shift of an electronic state corresponding to diagram (b) of Fig. 2 can be written as:

$$\Delta\omega_{DW}(T)=\sum_{n,\mathbf{q}}\frac{M_{DW}(n,\mathbf{q})}{\Omega(n,\mathbf{q})M}(2n_B+1) \tag{7}$$

where $M_{DW}$ is the matrix element of the electron-two-phonon interaction, $\mathbf{q}$ runs over all phonon wavevectors in the Brillouin zone (BZ) and $n$ over the phonon branches. Under the assumption that $M_{DW}$ is nearly the same for all phonons, which is justified except at low frequencies, $\Delta\omega_{DW}(T)$ is proportional to $\langle u^2\rangle$. In a more general way it holds that $\Delta\omega_{DW}(T)$ is proportional to $T$ for $T\gg T_D$ and to $M^{-\frac{1}{2}}$ for $T\ll T_D$.

The diagram of Fig. 2(a) corresponds to the first-order electron-phonon interaction treated in second order perturbation theory. It is a so-called self-energy diagram and represents a correction to the electron energy which has a real $(\Sigma_r)$ and an imaginary $(\Sigma_i)$ part (the complex self-energy $\Sigma$), it was first mentioned in the literature by H.Y. Fan [22] and is therefore often called the "Fan Term". Its analytical expression is similar to Eq. (7) except that $M_{DW}$ must be replaced by $|M_F|^2$, where $M_F$ is the matrix element of the first-order electron-phonon interaction, and an additional energy denominator must be introduced which has the form

$$\omega - \omega_\ell \pm \Omega - i\delta, \tag{8}$$

where $\omega_\ell$ represents all possible electronic intermediate states and ± absorption or emission of a phonon. $\delta$ is a positive infinitesimal. Wavevector conservation, but not energy conservation, is implicit in the excitations involved in Fig.2(a). The complication in the evaluation of the diagrams of Fig. 2(a) stems from the fact that the contributions of the individual phonons must be summed over all phonons and then must be summed over all possible intermediate states. If there are no intermediate states for which the real part of (8) vanishes, $\delta$ can be neglected and the corresponding self-energy turns out to be real. This happens above the highest valence band and below the lowest conduction band states. In the middle of the allowed bands there are states for which the real part of (8) vanishes and the corresponding self-energy has an imaginary part $\Sigma_i$ determined by a sum of delta functions, i.e., by the density of intermediate states $\omega_\ell$ [18]. Notice that $\Sigma_i$ has been defined to be negative. In this manner, the perturbed wavefunction of an electronic state decays with time, as corresponds to a lossy material. Away from the band edges, the imaginary parts of the self-energy which arise from Eq. (8) soon become much larger than the phonon frequencies and $\Omega$ can be neglected in the evaluation of $\Sigma$. At the conduction band edge, however, $\omega_\ell > \omega$ and only terms with +$\Omega$ in Eq. (8), corresponding to phonon absorption, generate a (small) contribution to $\Sigma_i$ (at finite temperatures). As soon as $\omega - \omega_\ell > \Omega$, i.e., slightly above the edge, the electron lifetime generated by electron-phonon interaction becomes finite even for $T$=0 because of the possibility of transitions with the emission of a phonon. Under conditions similar to those discussed in connection with Fig. 2(b), but somewhat more stringent because of the poles implied by Eq. (8), $\Sigma$ can also be assumed to be proportional to $\langle u^2 \rangle$. In any case, the proportionality to $T$ for $T \gg T_D$ and to $M^{-\frac{1}{2}}$ for $T \ll T_D$ also holds.

The discussion above applies to excited electrons in the conduction band as well as to holes in the valence bands: $\Sigma_i$ is also negative if one considers a hole in the valence band, but must be taken as positive for the corresponding electron (the hole amplitude decreases with time, whereas the corresponding electron amplitude increases as the empty electronic state is being filled). Many of the considerations above apply to electrons and holes as well as to the corresponding two-particle (electron-hole) excitations. The imaginary part of the self-energy of such an excitation is the sum of the corresponding electron and hole self-energies and, with our definition, remains negative (i.e., in an *optically passive* material the excitation decays with *time* through absorption and emission of phonons). The sign of the real energy correction of Fig.

2(b) is usually, but not necessarily, negative, i.e., the gap decreases with increasing temperature and with decreasing isotopic mass. This is so because the gap is opened, in a free-electron scheme, by the pseudopotential repulsion which becomes smaller with increasing $T$ on account of the Debye-Waller screening. The self-energy terms of Fig. 2(a), however, can lead to either positive or negative contributions to the gap energy because the energy denominator can have either sign.

Let us consider, e.g., the lowest gap. The corresponding conduction band state can interact with intermediate states in either the conduction or the valence band. The former interactions result in a negative contribution to the gap energy because the real part of Eq. (8) is negative. The latter result in a positive contribution. The same conclusion can be reached for the topmost valence band electron. Hence, considerable cancellation is expected. Because of the fact that the energy denominators of "intraband" terms are smaller than those of "interband" terms, however, the former will usually dominate. In this case, a negative contribution to the gap energy obtains. However, some cases are known in which gaps increase with increasing temperature. E.g., for the lead chalcogenides [23,24] both the thermal expansion and the Debye-Waller contributions are positive (the individual sign of the "Fan terms" contribution is not presently known).

## 3. The thermal expansion and contribution of zero-point vibrations to the volume at $p = 0$.

The thermal expansion (usually included in the quasiharmonic approximation) is an effect of the anharmonicity of the interatomic potentials. In the case under consideration, simple thermodynamics leads to the expression [25,26]:

$$\left(\frac{\Delta V(T)}{V_0}\right)_p = \frac{1}{2B_0 V_0} \sum_{n,\mathbf{q}} \hbar \Omega_{n,\mathbf{q}} \gamma_{n,\mathbf{q}} (2n_B + 1), \tag{9}$$

where $\gamma_n(\mathbf{q}) = -\left(d\ell n \Omega_n(\mathbf{q})/d\ell n V\right)_T$ are the so-called mode Grüneisen parameters, which can be either obtained from lattice dynamical calculations or experimentally determined under pressure, and $B_0$ is the bulk modulus.

For $T \gg T_D$ Eq. (9) can be written as:

$$\left(\frac{V(T)-V(0)}{V_0}\right)_p = \frac{\gamma_G C_v}{B} T \tag{10}$$

where $C_V$ is the heat capacity per unit volume and $\gamma_G$, first defined by Grüneisen [27], is the average mode Grüneisen parameter. Equation (10) indicates that the contribution of the thermal expansion to the temperature dependence of the gap [Eq. (1)] is linear in $T$ for $T \gg T_D$, like that of the anharmonic terms of Fig. 2.

For $T \ll T_D$ Eq. 9 can be written as:

$$\left(\frac{V(T)-V(0)}{V_0}\right)_p = \frac{\gamma_G' C_v}{4B} T \tag{11}$$

where $\gamma_G^{'}$ is an average of the Grüneisen parameters of the three acoustic phonon branches [28].

Using the Debye expression $C_V \propto (T/T_P)^3$ one finds with Eq.(11):

$$\left(\frac{\Delta V(T)}{V_0}\right)_p \propto \gamma_G^{'} M^{3/2} T^4 . \qquad (12)$$

The $T^4$ dependence also applies to the sum of the two contributions to gap shifts shown in Fig. 2 [29] (see Sect. 9).

Equations (10),(11) do not include the zero-point anharmonic renormalization which is represented by the summand "1" in the statistical factor $(2n_B+1)$ of Eq. (9). This renormalization $\frac{\Delta V_0}{V_0}$ is given by:

$$\left(\frac{\Delta V_0}{V_0}\right)_P = \frac{1}{2B_0 V_0} \sum_{n,\mathbf{q}} \hbar \Omega_n(\mathbf{q}) \gamma_{n,\mathbf{q}} \propto M^{-1/2} \qquad (13)$$

The $M^{-1/2}$ dependence of the zero-point renormalization is only valid for elemental crystals. For crystals with more than one kind of atom, the dependence on isotopic mass may be different for each constituent atom [26].

Equation (13) can be used in connection with Eq. (1) to determine the contribution of the "zero point thermal expansion" to the gap renormalization. Equations (10-13) depend on averages of $\gamma_{n,\mathbf{q}}$ over the BZ. Usually, $\gamma_{n,\mathbf{q}}$ does not vary much throughout the BZ (typical values lie between +1 and +2) except for the TA phonons for which, in tetrahedral semiconductors, it reaches negative values at the edge of the BZ (exceptions are diamond and possibly boron nitride) [30]. At temperatures close to that of the TA phonons, the negative Grüneisen parameters contribute a significant negative term to Eq. (9) which strongly lowers the values of $(\Delta V(T)/V_0)_p$. This contribution reverses the sign of $(\Delta V(T)/V_0)_p$ and the thermal "expansion" (with respect to $T = 0$) becomes a contraction. Close to the center of the BZ, the $\gamma$ of the TA phonons becomes positive again for Ge, Si, and the III-V compounds (but not for the II-VI compounds) [26,30] and thermal *expansion* obtains again at the lowest temperatures, at which Eq. (12) becomes valid. Because of the oscillations in $\gamma_{n,\mathbf{q}}$ just described, thermal expansion is very small at low *T*. A schematic diagram of the phenomena just described, and their universality, can be seen in Fig. 6 of [31]. The expansion coefficient of Ge, Si, and most III-V compounds shows two sign reversals, one for $T\sim 0.2\ T_D$ and the other at $T\sim 0.04\ T_D$. That of the II-VI compounds shows only one at $T\sim 0.2\ T_D$ whereas that of diamond, SiC, and possibly BN, shows no sign reversal at all.

Except at very low temperatures, it is possible to approximate the phonon dispersion relations with a single "Einstein" oscillator in order to evaluate the temperature dependence of $(\Delta V(T)/V_0)_p$ with Eq. (9). The corresponding fit to experimental data [32] for $(\Delta a(T)/a_0)_p$ (where $\Delta a(T)$ is the linear expansion) is shown in Fig. 3 for silicon. The fitted value of the Einstein frequency is $\Omega_E \approx 570$ cm$^{-1}$ ($T_E = 838$ K), slightly larger than the maximum frequency $\Omega \approx 525$ cm$^{-1}$. Within the experimental range in Fig. 3 (T >100 K) the sign reversals mentioned above do not affect $\left(\Delta a(T)/a_0\right)_p$. They can be seen, however, in the *T*-derivative of this function, i.e., the thermal expansion coefficient α(*T*), which reverses sign below $T \approx 120$ K. Note in Fig. 3 the linear *T*-dependence of $\left(\Delta a(T)/a_0\right)_p$ at high *T*, which was already surmised from Eqs. (9) and (10). The linear asymptote can be obtained rather accurately from the Bose-Einstein oscillator fit (dashed line in Fig. 3). It is easy to see, using Eq. (9), that the extrapolation of the straight line to *T* = 0, i.e., the *y*-axis intercept of the dashed line, determines the unrenormalized lattice parameter whose difference to $\left(\Delta a(0)/a_0\right)_p$ determines the zero-point renormalization $\left(\Delta a/a_0\right)_{p=0} =$ +1.86 x 10$^{-3}$. This renormalization can also be estimated by extrapolating to $M = \infty$ the dependence of *a*(0) on isotopic mass, which should be proportional to $M^{-\frac{1}{2}}$ (Eq. (13)). $\left[d\ell n\, a(0)/dM\right]_p$ has been measured to be –3.2 x 10$^{-5}$ (amu)$^{-1}$ [33]. From this value, using the $M^{-\frac{1}{2}}$ dependence, we find the zero-point renormalization:

$$\left(\frac{\Delta a}{a_0}\right) = +2M \times 3.2 \times 10^{-5} = +1.85 \times 10^{-3}, \quad (14)$$

where we have taken for *M* the average isotopic mass *M* = 29. The agreement between this result and that obtained from the linear extrapolation in Fig. 3 (+1.86 x 10$^{-3}$) is excellent. Low temperature anomalies in $\left(d\ell n\, a(0)/dM\right)_p$, related to the sign reversals of the thermal expansion discussed above, have also been recently observed [20].

Figure 4 displays the isotopic effect on the lattice parameter of germanium for $^{73}$Ge, $^{74}$Ge, and $^{76}$Ge, measured with respect to $^{70}$Ge $\left[Y = \{a(M)-a(70)\}/a(70)\right]$ in the 0 – 300 K range. As expected, the effect is largest (*Y* = -4.7x10$^{-5}$ for M = 76) for T→ 0 and decreases considerably with increasing *T* (it should tend to zero for *T*»*T*$_D$ = 374 K). The curves through the measured points were obtained with the following single Einstein-oscillator expression:

$$Y = 3.1 \times 10^{-5} (M-70)\left[\frac{1}{e^\beta -1}\left(-1+\frac{\beta}{1-e^{-\beta}}\right)-\frac{1}{2}\right] \quad (15)$$

with β = $T_D/T$ ($T_D \approx 374$ K). Equation (15), in which we only used the prefactor as an adjustable parameter, represents the experimental data rather well.

Before closing this section, we mention extant measurements and calculations for diamond (25) which yield $\frac{a(M=13)-a(M=12)}{a(M=12.5)} = 1.5 \times 10^{-4}$. It has been pointed out that because the lattice parameter of $^{13}$C diamond is smaller than that of $^{12}$C diamond, the chemical bonding of the former, and correspondingly its hardness, should be stronger [35]. This effect on the hardness, however, is too small to be of practical significance.

## 4. Absorption edges of germanium and silicon

The lowest absorption edge of both Ge and Si is indirect. For Ge, it involves transitions between the valence bands at the Γ point of the BZ and the conduction bands at L. For Si the indirect transitions take place between Γ and Δ (close to X). The different location in ***k***-space leads to the fact that the thermal expansion effect has the same sign as the electron-phonon interaction for Ge but the opposite one for Si. In the case of Si, the expansion effect is rather small because of the small value of a (see Eq.1) [29]. The temperature dependence of the indirect gaps of Ge and Si has been known for half a century [36]. Semiempirical analytic expressions for this dependence have been proposed, starting with the rather popular (but incorrect at low $T$ [29]) Varshni expression [37].

$$\omega = \omega_0 - \alpha T^2 / (T + \beta). \tag{16}$$

Einstein oscillator expressions have also been fairly successfully used, although they necessarily break down at low $T$ [for the same reason as they breakdown in the case of $C_v(T)$] [11]. We have already discussed Einstein oscillator fits in connection with the thermal expansion (see Fig. 3): The fitted oscillator frequencies are close to $T_D$. This fact also holds for the electron-phonon contributions to the dependence of gaps on $T$. It is therefore reasonable (at least for group IV materials, see Sect. 7 for other cases) to perform single Einstein oscillator fits to the total dependence of a given gap on $T$. Two-oscillator models, in the manner proposed by Lindemann and Nernst to interpret $C_v(T)$ [38], have also been used [39]. The two-oscillators roughly correspond to acoustic and optic phonons, respectively. A number of alternative empirical expressions which reproduce well the low-temperature $\left(\sim T^4\right)$ as well as the high temperature $(\sim T)$ behavior, at the expense of a larger number of adjustable parameters, have also been used [40]. The zero-point renormalization of electronic gaps, and the concomitant dependence on isotopic mass, however, have been only considered during the past decade [11,41,42,43].

We shall compare next the zero-point renormalization of these indirect gaps, as obtained from the extrapolation of their temperature dependence to $T = 0$, to the values obtained from the corresponding dependence on isotopic mass. In the case of germanium, not only the Γ→L indirect gap, but also the Γ→ Γ direct gap and its spin-orbit split component are accessible to experiments [41].

Figure 5 shows the indirect gap of silicon vs. $T$. The points are experimental [44,45]. The curve drawn through them represents a fit with the empirical expression [11,40]:

$$\omega_{ind}(T) = \omega_{ind}(0) - \frac{\zeta T_D}{2}\left[\left\{1 + \left(\frac{2T}{T_D}\right)^p\right\}^{1/p} - 1\right]. \tag{17}$$

The parameters in Eq. (17) have the values $\omega_{ind}(0) = 1.170$ eV, $\zeta = 0.318$ meV·K$^{-1}$, $p = 2.33$. A zero-point renormalization $-\zeta T_D / 2 = -64$ meV is obtained by extrapolating the asymptotic behavior found from Eq. (17) for $T \to \infty$ (see dashed line in Fig. 5) to $T = 0$.

It is important to note that the experimental *T*-range of Fig. 5 (confined to $T<T_D$) does not suffice to determine the asymptotic behavior for $T\rightarrow\infty$. A fit to the analytic expression is needed: a much smaller value for the gap renormalization would be found if one simply "eyeballs" the asymptote following the experimental points. The experimental range is usually limited (to 300 K in Fig. 5) by the broadening of optical spectra with increasing *T*. This problem does not arise when fitting thermal expansion data which, in Fig. 3, extend to 1500 K $\gg T_D$. The asymptote to the measured data can then be "eyeballed" rather accurately.

As already mentioned, the zero-point renormalization of a gap can also be estimated from the corresponding isotope effect. In the case of $\omega_{ind}$ for silicon, a gap shift of –1.04 meV/amu is found for $T \gg T_D$. On the basis of the $M^{-1/2}$ dependence of the gap renormalization, the isotope effect leads to a renormalization of 60±1 meV for the $\omega_{ind}$ of silicon, in reasonable agreement with the value found from the $\omega_{ind}(T)$ asymptote.

The isotopic shift, and the corresponding zero-point renormalization, have also been determined for the indirect gap of Ge (see [41] and references therein). They amount to (at $T \ll 6$ K): isotope shift = 0.36 meV/amu, renormalization = −53 meV. The Pässler fit to $\omega_{ind}(T)$ yields −52 meV for the renormalization [40], also in good agreement with the value obtained from the isotope effect. For the direct gap of germanium $\omega_0$, a value of 0.49 meV/amu is reported in [41]. The corresponding gap renormalization (obtained by using the $M^{-1/2}$ law) is –71 meV, whereas a value of −60 meV is obtained by "eyeballing" the linear asymptote of $\omega_0(T)$ in Fig. 6.44 of [14]. An average renormalization value of –62 meV has been calculated by LCAO techniques in [17]. This gap exhibits a spin-orbit splitting $\Delta_0$ and the corresponding isotope effect has been measured for both components, $E_0$ and $E_0+\Delta_0$ [41]. Surprisingly, the effect on the $E_0+\Delta_0$ gap of germanium has been found to be 30% larger than that on $E_0$, a fact which cannot be explained on the basis of our present understanding of the spin-orbit interaction. Actually, the authors of [34] have recently found [43] that the spin-orbit split component of the $\omega_{ind}$ of Si shows the same isotope effects as the main component, a fact which suggests that the isotope effect measurements of $E_0+\Delta_0$ presented in [41] should be repeated.

**6. Diamond**
*6.1 Dependence of the edge luminescence frequency on temperature and isotopic mass*

The luminescence at the indirect exciton of diamond (which is silicon-like) is shown in Fig. 6 [46]. The curve through the experimental points corresponds to a single Einstein oscillator. The fitted frequency, 1080 cm$^{-1}$ = 1580 K, is close to the Debye temperature (~1900 K). This figure illustrates again the fact mentioned in connection with Fig. 5: in order to obtain the asymptotic behavior of a gap for $T\rightarrow\infty$ we must either have data for $T > T_D$ (not the case in Fig. 6) or fit the available data to a reliable algebraic expression, as has been done for Fig. 6. The thick straight line depicts the corresponding asymptote which enables us to estimate a gap renormalization of 370 meV, much larger than the corresponding values for Ge and Si ($\approx 70$ meV). In order to corroborate this *a priori* unexpected result, data on the corresponding isotope shift come in handy. There are two stable carbon isotopes, $^{12}$C and $^{13}$C. In [47] we find a derivative of the gap with respect to $M$ equal to to $14\pm 0.8$ meV/amu which, using the $M^{-1/2}$ rule, leads to the renormalization $-2 \times 14 \times 13 = -364$ meV, in excellent agreement with the value estimated above. A recent semiempirical LCAO calculation results in a renormalization of 600 meV [48], even larger than the experimental values. The value of the exciton energy shown in Fig. 6, 5.79 eV, can be compared with *ab initio* calculations of the indirect gap [49], which yield an indirect gap of 5.76 eV (average value in Table III of [49]). This value must be compared with that from Fig. 6 after adding the exciton binding energy (5.79+0.07=5.86 eV). This experimental gap value is in rather good agreement with the *ab initio* calculations. The agreement becomes considerably worse if the zero-point renormalization of 370 meV is not taken into account (5.48 vs. 5.76). All these arguments support the large zero-point renormalization found for the indirect gap of diamond. Using Eq.(1), the corresponding volume deformation potential $a = -2.2$ eV and the volume renormalization $\Delta V/V = 3.8\times 10^{-3}$ (Sect. 3), we find the quasiharmonic effect on the gap to be 8.4 meV (see also Fig. 4 of [47]). This is about 2% of the total renormalization of Fig. 6 which is therefore dominated by the electron-phonon diagrams of Fig. 2.

The next question is why the gap renormalization is so large for diamond. It has been suggested that other materials with elements of the second row of the periodic table (e.g., ZnO, GaN, possibly SiC) also exhibit strong electron-phonon renormalization effects [50]. A simple, hand-waving explanation can be found in the fact that C (and also N,O) does not have core *p*-electrons. Hence, the vibrations of these atoms affect strongly their $2p$ valence electrons which see the full unscreened core potentials. In order to check this conjecture, it is interesting to compare the valence band deformation potentials of diamond with those of Si and Ge. The deformation potential $d_0$ which describes the splitting of the valence bands of these materials induced by a frozen optical phonon of $\Gamma_{25'}$ symmetry [51]. $d_0 \cong 63$ eV has been calculated for diamond and 30 eV for Si (semiempirical calculations give values of $d_0$ as large as 100 eV for diamond [51]). Effects of the type of Fig. 2(a) related to the $\Gamma_{25'}$ phonons are thus expected to be 4.4 times stronger in diamond than in Si. This fact is likely to be relevant to the superconductivity of boron-doped diamond which will be discussed next.

*6.2 Superconductivity in boron-doped diamond*

Superconductivity with $T_c$ up to nearly 9K has been recently discovered in heavily boron-doped diamond prepared in two different ways at different laboratories [52,53]. The samples investigated in [52] showed superconductivity for $T<$ 4K $(T_c \square 4K)$. They were produced by the high-pressure high-temperature technique. The samples used by Takano *et al*. [53] were thin films prepared by microwave-plasma-assisted chemical-vapor-deposition and reached $T_c$'s close to 9K (see Fig.7). Although the hole concentration $n_h$ was not exactly known, for the present discussion we shall assume it to be close to the boron concentration of $2\times 10^{21}$ reported in [52] (i.e., 1.1% of the carbon atoms were substituted by boron). For these values of $n_h$, the holes in diamond should be degenerate, i.e., the sample should be metallic-like.

Because of the simplicity of degenerate semiconductors and the strong hole-phonon interaction mentioned in the previous section, it is natural to conjecture that the superconductivity discovered in B-doped diamond is BCS-like, with pairing induced by hole-phonon interaction. Two more or less elaborate calculations of $T_c$ for B-doped diamond are available in preprint form [54,55]. Because of the uncertainties involved in these calculations, we shall give here the simplest possible version of the corresponding theory in order to clearly illustrate the origin of the phenomenon and indicate the effect of the various parameters on the estimate of $T_c$. Isotope effect measurements, of the essence to confirm the BCS nature of the phenomenon, have yet to be performed.

The basis of these BCS-like calculations of $T_c$ is the equation [56]:

$$T_c = T_D \exp\left\{-\left[\lambda(1+\lambda)^{-1} - \mu^*\right]^{-1}\right\}, \tag{17}$$

where $\lambda$ is the dimensionless hole-phonon interaction parameter and $\mu^*$ represents the Coulomb repulsion which, because of the uncertainties involved in $\lambda$, we will set equal to zero. For the Debye temperature we shall take $T_D = 1900$ K (see Sect. 6a). It thus remains to estimate $\lambda$ in order to obtain an estimate of $T_c$ with Eq. 17. The value of $\lambda$ is given by [56]:

$$\lambda = \frac{N_d D^2}{M \omega_D^2} \tag{18}$$

where $N_d$ is the density of states (per spin and primitive cell, PC) at the Fermi level $E_F$, $M$ the atomic mass (12 amu) and $\omega_D$ the Debye frequency equivalent to 1900 K (1300 cm$^{-1}$ = 161 meV). $D$ is an electron-phonon interaction parameter (per unit phonon displacement) averaged over the Fermi surface and including all phonons. We start our task by evaluating $E_F$ and $N_d$. We neglect the spin-orbit interaction which is small in diamond (~15 meV at the $\Gamma$-point of the upper valence bands [57]). Three degenerate warped bands meet at this point. Their longitudinal masses, which depend on the direction of $k$, are given in Table II of [58]. The total density of states at $E_F$, obtained by adding the contributions of the three bands, turns out to be $N_d = 0.047$ states/(spin, eV, PC) = 1.3 states/(spin, Hartree, PC). We have given here also the density of states in Hartree (H), since we shall evaluate (18) in atomic units: $\omega_D = 5.9 \times 10^{-3}$ H and M = $12 \times 1830 = 2.2 \times 10^4$ electron masses (a.u.). The most difficult part in the evaluation of Eq. (18) concerns the average hole-phonon interaction parameter $D$. We start with the deformation potential $d_0$ mentioned in the previous section, for which we will take the value of 85 eV = 3.13 H, which falls in the upper range of the various estimates given in Table 1 of [51] (from 54 to 106 eV). This deformation potential, when multiplied by $(u/a_0)$, yields the energy shift of the light-hole band along [111] for a $\Gamma_{25'}$ phonon corresponding to an expansion of the [111] bond by an amount $u$. Tedious averaging over Fermi surface and phonons can be simplified by noticing that the hole effective mass is by far the largest for the top valence band and $k$ along {110}[58]. Hence, the regions of this band around {110} give the largest contribution to $N_d$ and we can confine ourselves to the evaluation of the average hole-phonon interaction at these points, including "intravalley" ( [110] to [110]) and "intervalley" ( e.g., [110] to [011]) hole-phonon interaction. The corresponding average matrix element for the phonon displacement $u$ is $(3/4a_0)d_0 u = 0.33u$ H ($u$ in Bohr). Following this procedure we find $\lambda = 0.20$ for $d_0 = 85$ eV. Replacing all these parameters into Eq. 17 we obtain $T_c = 5.2$ K. If we use instead $d_0 = 80$ eV, which is closer to the average of the values in Table 1 of [51], we find $T_c = 2.71$ K. These two values of $T_c$ bracket the experimental value reported in [52]. The large difference in the $T_c$ for two values of $d_o$ differing only by 6% illustrates the uncertainties in the calculated values of $T_c$ which can thus only be regarded as heuristic.

The question now arises of whether superconductivity could also be observed in *p*-type Si and/or Ge. Dopings of $n_h = 2 \times 10^{22}$ cm$^{-3}$ are surely difficult but not impossible to reach in these semiconductors. Taking $d_0 = 30$ eV and the hole masses given in [14] we find for this value of $n_h$, $\lambda = 0.03$ which for $\mu^* = 0$ would lead to a negligibly small $T_c$. A very small value of $\mu^*$ ($\mu^* \geq 0.03$) would suffice to suppress the superconductivity in Si and also in Ge, even at the lowest conceivable temperatures. Thus, the superconductivity observed in diamond is to be regarded as resulting from the strong hole-phonon interaction and the large density of states mass. We shall see in the next section that ZnO and GaN also have a strong hole-phonon interaction (although half as large as that of diamond). If they could be acceptor-doped to the levels reached in diamond they may be likely candidates for the observation of superconductivity. Such doping levels seem to be, at present, rather difficult to reach.

*6.3. Dependence of the refractive index on temperature and on isotopic mass*

The extrapolation of the high-temperature asymptote to $T = 0$ has been shown, in the previous sections, to be a powerful technique for determining zero-point renormalizations of temperature-dependent properties. It has also been used to estimate the corresponding anharmonic renormalizations of the elastic constants [11] and the phonon frequencies [59]. For the Raman frequency of diamond, an anharmonic zero-point softening of 20 cm$^{-1}$ is found [59]. An estimate based on the isotope effect measured for this frequency leads to $16 \pm 4$ cm$^{-1}$ [59] whereas an *ab initio* calculation [60] yields 32 cm$^{-1}$. Einstein oscillator fits can also be used to represent the temperature dependence of the long wavelength refractive index $n(T)$ (i.e., in the region of transparency) of crystals. Figure 8 shows such a fit performed for diamond [61]. The Einstein oscillator frequency obtained from the fit is 711 cm$^{-1}$ (~1040 K), considerably smaller than the Debye frequency $(\omega_D \square 1300 \text{cm}^{-1})$. Actually, 711 cm$^{-1}$ corresponds to the center of the TA phonon density of states band. This suggests that these phonons provide the main contribution to the temperature dependence of $n(T)$.

From the fitted Bose-Einstein expression (thick solid curve in Fig. 8) and its corresponding $T \to \infty$ asymptote (thin straight line), one obtains a zero-point renormalization $\Delta n(T=0) = +0.010(4)$. Using the by now well known $M^{-1/2}$ dependence, we estimate for the difference between $\Delta n(T=0)$ between $M = 13$ and $M = 12$, $[\Delta n(12) - \Delta n(13)] = \frac{1}{2} \frac{0.010}{12.5} = 4 \times 10^{-4}$, a value nearly a factor of three smaller than that *ab initio* calculated in [62]. A determination of this isotope effect using two diamond wafers with different isotopic compositions would therefore be of interest.

## 7. Binary Compounds
*7.1 Gallium nitride, zinc oxide, cadmium sulfide*

We have already mentioned the possibility of using *two* Einstein oscillators (Lindemann's model [38]) for representing the temperature dependence of physical properties, in particular electronic energy gaps [39,40]. The need for two oscillators becomes particularly acute for binary compounds in which cation and anion have very different masses (GaN [50], ZnO [63], CdS [65], CuI [64]). In these cases the acoustic and the optical phonons have rather different average frequencies and, in view of their different coupling to electronic states, it is not possible to replace them by a single

Einstein oscillator in order to describe the temperature as well as the isotopic mass dependence of energy gaps. Two frequencies, however, usually do the job (in the case of CuI three frequencies are needed [64]). The expression used for the two-oscillator fit of gaps is:

$$\omega_0(T) = \omega_0 - 2M_c \frac{d\omega_0}{dM_c}[2n_B(\Omega_c/T)+1] - 2M_a \frac{d\omega_0}{dM_a}[2n_B(\Omega_a/T)+1], \quad (19)$$

where the subscripts $c$ and $a$ represent the cation and the cation, respectively. We shall discuss here explicitly the case of the very topical material gallium nitride ($c$ = Ga, $a$ = N) which usually crystallizes in the wurtzite structure (zincblende-type crystals of GaN exist but are difficult to prepare).

A number of choices of adjustable parameters are possible in order to fit experimental data for $\omega_0(T)$ with Eq. (19). The four disposable parameters are $d\omega_0/dM_c$, $d\omega_0/dM_a$, $\Omega_c$ and $\Omega_a$. $\omega_0$ can often be eyeballed (at least to a first approximation) through the asymptote construction. Four parameters may still be too many when it comes to fitting monotonic and not very structured curves (as we will see in the next subsection, $\omega_0(T)$ is nonmonotonic for the copper halides and more parameters can be extracted from the fit). If the isotope effect coefficients are known, $\Omega_c$ and $\Omega_a$ are the only disposable parameters available and their fitted values must fall within the rather narrow ranges of acoustic and optic phonons averaging at 226 cm$^{-1}$ (330 K, acoustic) and 650 cm$^{-1}$ (950 K, optic) [50]. Unfortunately, no samples of $^{69}$GaN and $^{71}$GaN are available until now, although Ga$^{14}$N and Ga$^{15}$N thin films have been grown and measured by photo- and cathodoluminescence [50]. Using the measured isotope coefficient $d\omega_0/dM_N$ = 5±1 meV/amu and adjusting the remaining three free parameters we obtain the fit to the experimental data shown in Fig. 9 with the fitted values $d\omega_0/dM_{Ga}$ = 0.39 meV/amu, $\Omega_{Ga}$ = 327 K and $\Omega_N$ = 975 K. Note that $\Omega_{Ga}$ and $\Omega_N$ = are rather close to the average peak values of acoustic and optic densities of phonon states mentioned above. From Eq. (19) we also obtain the zero-point renormalization:

$$\omega_0(0) - \omega_0 = -2M_c \frac{d\omega_0}{dM_c} - 2M_a \frac{d\omega_a}{dM_a} = -200 \text{ meV} \quad (20)$$

This renormalization is very large compared with that of the corresponding direct gap of germanium (-53 meV, Sect. 5). It agrees, however, with the average renormalization of the direct gap of Ge and the indirect of diamond (no data for the direct gap of diamond are available): -(53+364)/2 = 208 meV. This is an interesting observation: in GaN Ga is the periodic table neighbor of Ge and N that of carbon. We thus surmise that for elements of the carbon row in tetrahedral structures the zero-point gap renormalizations are large. This fact has already been attributed to the lack of core *p* electrons in these elements, a conjecture that is corroborated by a similar investigation performed for ZnO [63].

*7.3 Cuprous halides*

For all materials and gaps discussed so far, an increase in the isotopic mass results in an increase in the gap because of the corresponding decrease in the *negative* zero-point renormalization. The cuprous halides, CuCl, CuBr, and CuI, isoelectronic and isostructural (zincblende) to many of the materials just discussed, are anomalous in this respect. An *increase* of the copper mass results in a *decrease* in the gap ($d\omega_0/dM_c < 0$) [64]. This effects a reduction of the total zero-point renormalization of the gap [11,64]. Similar anomalies seem to occur also in the Cu- and Ag-based chalcopyrites (I-III-VI$_2$ compounds, e.g., CuGaS$_2$, AgInSe$_2$) [64].

The isotopic effects of Cu, Cl and Br have been measured in the copper halides [11,64] (see Fig. 10 for CuI). Unfortunately, iodine has only one stable isotope ($^{127}$I) and therefore the mass effect of the anion cannot be measured directly. It can, however, be inferred to be $d\omega_0/dM_I = +0.24$ meV/amu from a *three oscillator* fit to the rather unusual and nonmonotonic temperature dependence of the edge exciton energies shown by the dots in Fig. 11. The small temperature shift observed results from a cancellation of cation and anion contributions.

In order to fit the rather structured remainder of this cancellation three oscillators, corresponding to TA, LA, and LO phonons, are needed (see arrows in Fig. 11).

## 8. Temperature and Isotope Effects on the Full $\varepsilon_2(\omega)$ Spectra and the Corresponding Interband and Critical Points

We mentioned in connection with Fig. 1 that state-of-the-art computational techniques allow the calculation of $\varepsilon_2(\omega)$ spectra but not their temperature dependence. They do not even allow the calculation of $\varepsilon_2(\omega)$ for $T = 0$ since they assume atoms at fixed crystallographic positions, implicitly neglecting zero-point vibrations. Unfortunately, theorists often do not even bother to compare their calculations with available measurements for $T \ll T_D$, using instead more easily accessible room temperature spectra. This is illustrated in Fig. 12 where we show calculations with and without electron-phonon interaction, as compared with measurements at 300 K and 20 K [9,11,66].

The most striking fact in this figure is the strong effect of the excitonic interaction: a strengthening of the $E_1$ structure seen around 3.5 eV (by nearly a factor of two) together with a weakening of the $E_2$ and $E_1'$ structures (4.5 and 6 eV) so as to preserve the sum rule which applies to $\omega \cdot \varepsilon_2(\omega)$. This effect is of a canonical nature: It is observed in most materials under consideration. The second interesting observation is the fact that the spectrum calculated including exciton interaction is much closer to the one measured at 20 K than to that measured at 300 K, especially around the $E_1$ peak (which is mainly excitonic). There is also a shift by about 90 meV between the spectrum calculated with exciton interaction and that measured at 20 K. Although this shift is probably within the error bars of the calculation, it agrees with the zero-point renormalization which was estimated in [67] using semiempirical pseudopotentials (electron phonon effect = −75 meV; thermal expansion = −25 meV; total zero-point renormalization = 100 meV). From the measured isotope effect, a zero-point renormalization of 118 meV is obtained for the $E_1$ transitions of silicon [68].

Since the states included in the $E_1$ transitions of Si are not at the fundamental gap, the electron-phonon interaction is expected to broaden them. This zero-point broadening, of course, should also be proportional to $M^{-1/2}$ and can, in principle, be estimated from the dependence of the linewidth of $E_1$ on $M$ [68]. The range of $M$'s available, however, does not suffice to obtain reliable values of the Lorentzian broadening $\Gamma_{E_1}$ which can be calculated to be (using semiempirical pseudopotentials) $\Gamma_{E_1}(T=0) = 47$ meV [67]. A similar value of $\Gamma_{E_1}(T=0)$ is obtained from the measured temperature dependence of $\Gamma_{E_1}$ using the Einstein oscillator fit $\Gamma_{E_1}(T) = \Gamma_{E_1}(T=0) \bullet \left[1 + 2/\left(e^{T_D/T} - 1\right)\right]$. The fit (see Fig. 13) yields the parameters $T_D = 743$ K and $\Gamma_{E_1}(T=0) = 59$ meV.

In the case of germanium, the $E_1$ transitions are split into $E_1$ and $E_1 + \Delta_1$ by spin-orbit interaction. The zero-point energy renormalization of these transitions has been found to be $-124 \pm 30$ meV for $E_1$ and $-170 \pm 60$ meV for $E_1 + \Delta_1$. The corresponding measured broadenings are $\Gamma_{E_1}(T=0) = 30 \pm 7$ meV and $\Gamma_{E_1+\Delta_1} = 45 \pm 7$ meV [13]. These values agree rather well with the pseudopotential calculations of [48].

## 9. Temperature Dependence of Energy Gaps at Very Low Temperatures ($T \ll T_D$)

*9.1 Indirect gap of silicon below 5K*

It has been mentioned in Sect. 3 that Einstein oscillator fits should not work at very low temperatures. Two-oscillator fits extend, of course, to lower temperatures than a one-oscillator fit [38], but they are still bound to deviate from experimental data, typically in the region $T < 0.01 T_D$. This fact led Peter Debye to introduce his theory of the specific heat based on the treatment of the acoustic phonons as quantized mechanical oscillations of a continuum. The result is a $T^4$ dependence for the total vibrational energy ($T^3$ for the specific heat), similar to that found for the thermal expansion (Eq. (12) and its contribution to the temperature dependence of gaps.

The contribution of the diagrams of Fig. 2 to the shift of gaps with temperature is also similar to that of Eq. (9), i.e., to the contribution of the thermal expansion when Eq. (9) is combined together with Eq. (1). One must, however, replace in Eq. (9) the term $\Omega_{n,q}$ by $|\mathbb{M}_{n,q}|^2/\Omega_{n,q}$, where $|\mathbb{M}|^2$, represents a coupling constant of the electron-phonon interaction which includes the terms (a) and (b) in Fig. 2. The contribution of Fig. 2b to $|\mathbb{M}_{n,q}|^2$ is straightforward. The corresponding contribution of Fig. 2a may be more complicated since it involves resonant energy denominators. We shall assume that the resonance produced by the energy denominator is negligible because of the small volume of **k**-space it encompasses, and shall treat the total $|\mathbb{M}_{n,q}|^2$ as a smooth function of **q**, a fact which is supported by the experimental results to be discussed next. Acoustic phonons become for $\mathbf{q} \to 0$ uniform translations of the lattice, a fact which requires that $|\mathbb{M}_{n,q}|^2$ be proportional to $\Omega_q^2$ to the lowest order in $\Omega_q$. This additional factor of $\Omega_q^2$ cancels that in the denominator of $|\mathbb{M}_{n,q}|^2/\Omega_q$ and leaves us with a factor of $\Omega_q$ inside of the summation, i.e., with an expression fully isomorphic to Eq. (9) as far as the dependence on $\Omega_q$ for acoustic phonons is concerned. This leads to a $T^4$ temperature dependence of the gap shift for $T < 0.01 T_D$ which, till the appearance of [29], had not been reported.

The observation of this $T^4$ dependence became possible after it was realized [69] that the zero-phonon luminescence lines of indirect excitons bound to impurities in silicon with the natural isotopic concentration are mainly broadened, at low $T$, by the isotopic mass disorder. The corresponding lines of isotopically pure (99.90%) $^{28}$Si become much sharper [69], possibly sharper than 0.15 µeV at 4K [29]. Under these conditions it became possible, using excitation spectroscopy with a tunable single mode Yb-doped fiber laser, to measure the temperature dependence of the peak frequency of the strongest boron-bound exciton line with an accuracy of 0.01 µeV. The results are shown in Fig. 14. They follow a $T^4$ dependence on $T$, but it was realized in [29] that this dependence is mostly due to the change in the equilibrium pressure of the helium bath resulting from pumping on it in order to set the temperature (see $\Delta E_p$ contribution in Fig. 14). This rather unusual effect enables the determination of the pressure dependence of the indirect gap using pressures below 1 bar! (note that the pressure and/or temperature dependence of the luminescence peak should closely follow that of the indirect gap). The equilibrium pressure of the helium bath also follows a $T^4$ law, a fact which can be understood on the basis of Eq. (12) and a bulk modulus for liquid He independent of temperature. Hence, the experimental points of Fig. 14 cannot be directly used in support of the $T^4$ temperature dependence of the energy gap at constant pressure. In order to remove the pressure effect from the experimental data, the authors of [29] lowered the temperature to just above the λ-point and then increased the pressure rapidly to that which corresponds to the highest $T$-point in Fig. 14. The increase in pressure was so fast that it did not give time for the temperature to rise while the luminescence was being measured. This revealed a pure pressure shift which led to the hydrostatic pressure coefficient of 1.52 µeV/bar, in agreement with data obtained using pressures up to $10^5$ kbar (10 Gpa) [29]. This coefficient can be used to eliminate the pressure effect from the experimental points of Fig. 14. In Fig. 15 the results of this procedure are shown to follow rather accurately the $T^4$ law.

*9.2. Other materials: $Cu_2O$*

Although the $T^4$ law has only been convincingly demonstrated for the indirect gap of isotopically pure silicon ($^{28}$Si), a few other indications can be found in the literature. Pässler, for instance, fitted the temperature dependence of the lowest gaps of 22 tetrahedral semiconductors with Eq. (17) which corresponds to the asymptotic behavior $T^p$ for $T \to 0$. He obtained values of $p$ between 2.0 and 3.3, averaging at $p = 2.5$. This value of $p$ lies below the $p = 4$ predicted here, a fact that is most likely due to having performed the fits over a broad temperature range, not only in the range $0 \leq 0.01\ T_D$: The data available to Pässler were not sufficiently accurate to perform a fit confined only to very low temperatures. It is interesting to note, however, that $p$ is larger for materials with large $T_D$ (diamond, $p = 3.3$; SiC, $p = 3.08$; AlN, $p = 3.0$) for which $T_D$ is large and the temperature range used for the measurements falls mostly in the $T \ll T_D$ region.

As the last item, we would like to discuss the temperature dependence of the gap of $Cu_2O$. Excellent crystals of this material can be found in nature as the mineral *cuprite*, although they can be also grown in the laboratory. The existence of such crystals plus the fact that there are very nice Rydberg-like excitonic series associated with the lowest absorption edge of $Cu_2O$, made it one of the most popular crystals for early investigations of excitonic spectra [71,72]. The crystal structure is simple cubic, with 2 molecules per primitive cell. There are therefore 18 phonon branches which fall into 3 rather well separated groups: one around 620 cm$^{-1}$, corresponding mainly to oxygen vibrations, and two around 100 and 230 cm$^{-1}$, corresponding probably to copper vibrations, with the four Cu atoms of the primitive cell undergoing vibrations of the bond-bending and bond-stretching type, respectively. Its absorption edge is direct but forbidden by parity. The corresponding absorption, and the oscillator strength of the so-called yellow edge excitons, is very weak. The first (n=1) exciton in the series is dipole-forbidden (but weakly quadrupole allowed). The corresponding n=2, 3 ... excitons are weakly dipole allowed. In [73] the energy of the n=1 exciton was measured vs. *T* in the $1 \leq T \leq 30$ K range. Figure 16 shows the shifts vs. *T* reported in [73] and a fit to a $T^4$ curve. Up to 20 K, the measured shifts follow well the $T^4$ dependence but fall a bit short of the $T^4$ prediction at 25 K and especially at 30 K, not surprisingly in view of the fact that for $Cu_2O$ $T_D \cong 200$ K.

Measurements of the temperature shift of the absorption edge related to the yellow exciton of $Cu_2O$ over a wider temperature range have been published earlier [74]. The experimental results are reproduced in Fig. 17. The curve in this figure represents a three oscillator fit. The frequencies of these three oscillators have been taken to be the average ones of the three groups of phonons mentioned above, corresponding to 140, 340 and 900 K, respectively. The number of adjustable parameters can be reduced to one by estimating the effect of the high frequency oscillators from the dependence of the zero-point renormalization on isotopic mass of oxygen reported in [75]: $d\omega_0/dM_o = 9$ cm$^{-1}$/amu. The coupling of the two copper-like oscillator groups can be estimated from recent measurements of the copper isotope effect on the n=3 yellow exciton which yield $d\omega_0/dM_{Cu} = 1.8$ cm$^{-1}$/amu. The contribution of the copper vibrations to the gap renormalization thus becomes: $-1.8 \times 2 \times 64 = -230$ cm$^{-1}$. We split this contribution into the contribution of the lower frequency oscillators, $-(230-\xi)$, and that of the intermediate frequency ones, $-\xi$. We thus obtain the following expression for the temperature dependence of the yellow exciton frequency.

$$\Delta \omega_0(T) = [\frac{230-\xi}{e^{140/T}-1} + \frac{\xi}{e^{340/T}-1} + \frac{306}{e^{900/T}-1}] \text{ cm}^{-1} \qquad (21)$$

Using $\xi$ as the only fit parameter we obtain the curve of Fig. 17 for $\xi = -190$ cm$^{-1}$. Note that the contribution of the 340 K oscillators to the gap renormalization has a sign opposite to either the total copper or the oxygen contributions. While this is in principle possible, considerable errors which may affect the measured isotope coefficients do not allow us to regard this conclusion as definitive.

## 10. Conclusions and outlook

We have discussed the dependence on temperature of the absorption spectra and energy gaps of many semiconductors and the corresponding zero-point (*T*=0) renormalizations which can be found from their temperature dependence and also from isotope effects. In binary semiconductors in particular, isotope effect measurements are of the essence for understanding the details of the electron-phonon interaction as applied to the temperature dependence of gaps. We have shown that for diamond (and also for GaN and ZnO) the zero-point renormalization of the lowest gap is particularly large (close to 400 meV for diamond). We have proposed that the correspondingly large hole-phonon interaction at the top of the valence bands is responsible for the BCS-type superconductivity recently observed in heavily boron-doped diamond. We have discussed recent measurements of the temperature dependence of the indirect gap of isotopically pure $^{28}$Si in the $1 \leq T \leq 5$ K region which enable us to confirm the $T^4$ dependence of this shift predicted theoretically.

It is hoped that measurements of gap energies will become possible in other isotopically pure semiconductors (*e.g.* Ge) with an accuracy close to that achieved for $^{28}$Si. This accuracy has enabled us to obtain values for the pressure coefficient of the gap in the pressure range below 1 bar. Uniaxial pressure experiments should be performed for $^{28}$Si (uniaxial stresses as high as 3Gpa are possible). Also, isotope experiments would be of considerable interest in the copper and silver chalcopyrites. We have pleaded with theorists to compare the properties they calculate, without considering electron-phonon effects, with experimental data obtained at low temperatures and, even better, with these data after electron-phonon renormalizations are removed. They will often (but, of course, not always) find that the comparison becomes more favorable after electron-phonon effects are properly taken into account. The goal of calculating full optical spectra $\varepsilon_2(\omega)$ including electron-phonon interaction should be pursued.

**Acknowledgments**

**Figure Captions**

Fig. 1 Temperature-dependent imaginary part of the dielectric function of germanium measured by ellipsometric techniques [12]. The spectrum starts at 1.4 eV and thus omits the rather weak ($\varepsilon_2 \sqcup 0.5$) $E_0$ and $E_0+\Delta_0$ direct transitions (~0.9 eV at 0 K). The even weaker indirect transitions (~0.74 eV at 0 K) are also outside of the spectral range. $E_1$, $E_1+\Delta_1$, $E'_0$, $E_2$ and $E'_1$ label interband critical points (van Hove singularities) in the standard notation [5,14].

Fig. 2 Electron-phonon interaction terms that contribute to the renormalization of electronic states. (a) Fan terms corresponding to a complex self-energy. (b) Debye-Waller terms leading to a real energy correction. The straight line represents an electronic state, the curly curve a phonon and the dots electron-phonon interaction vertices.

Fig. 3 Dependence on temperature of the linear thermal expansion of silicon $\Delta a_0/a_0$. The solid line represents X-ray data [32]. The points are a fit to these data using the standard Bose-Einstein expression given in the figure (with $T$ in Kelvin).

Fig. 4 Dependence of the lattice parameter of germanium $a(M)$ on isotopic mass $M$ plotted vs. temperature: $Y= [a(M)-a(70)]/a(70)$. The points are experimental, most of them from Ref. 34. The points labeled $S$ ($^{76}Ge$) are from Ref. 33. The solid curves were calculated with the single Einstein oscillator model of Eq.(15). See text.

Fig. 5 Temperature dependence of the indirect gap of silicon. The points are experimental [36], the solid curve represents a single Einstein oscillator fit to the experimental points. The dashed line represents the asymptotic behavior at high temperature: its intercept with the vertical axis allows us to estimate the bare gap and thus the zero-point renormalization due to electron-phonon interaction.

Fig. 6 Energy of the indirect exciton of diamond versus temperature. The points are experimental. The solid curve represents an Einstein oscillator fit whereas the dashed line represents the asymptotic behavior at high temperature as extracted from the Einstein oscillator fit. The intercept of the dashed line with the vertical axis determines the unrenormalized (bare) gap. See text.

Fig. 7 Resistivity of a boron-doped diamond film vs. temperature under several magnetic fields. The data indicate the onset of superconducting behavior with $T_c = 7$ K and full superconducting behavior for $T_c= 4.2$ K. From [53].

Fig. 8 Temperature dependence of the refractive index in diamond. The open circles represent the experimental data. The solid curve represents an Einstein oscillator fit to these data (see text). From [61].

Fig. 9 Energy of the free A exciton in GaN samples as a function of temperature up to 1100 K (H. Herr, diploma thesis, Technical Univ. of Munich, 1996). The black solid line corresponds to a fit of the data to the two-oscillator model of Eq. 19. From [50].

Fig. 10 Isotope mass dependence of the $\Gamma_3$ and $\Gamma_5$ edge excitons in CuI. The lines represent least-squares fits to the peak positions [64].

Fig. 11 Temperature dependence of the lowest exciton peak for CuI (equivalent to that of the $E_0$ gap). The dotted line represents a three-oscillator fit to the difference between the experimental data (solid circles) and the contribution due to the thermal expansion (open circles), which is displayed shifted in energy in order to show more clearly the relevance of this term. The dashed line is obtained by adding both contributions. The arrows indicate the frequencies of the three oscillators used in the fit. Note the flattening of the temperature dependence for $T \geq 150$ K [64].

Fig. 12 Comparison of *ab initio* calculated $\varepsilon_2(\omega)$ spectra of silicon (without electron-phonon interaction) with those measured at 20 K and 300 K. The dashed black curve represents calculations without excitonic interaction, whereas the continuous black curve represents calculations including this interaction (but no electron-phonon effects). From [9] and [11].

Fig. 13 Lorentzian broadening [half-width at half-maximum (HWHM)] of the $E_1$ excitonic peak of silicon vs. temperature, from [67].

Fig. 14 Frequency measured for the strongest zero-phonon luminescence line of $^{28}$Si vs. temperature. $\Delta E_p$ represents the effect induced by changes in the equilibrium pressure of the liquid He bath resulting from the changes in temperature. $\Delta E_T$ represents the true temperature effect. The points are well fitted by the $T^4$ law. From [29].

Fig. 15 Natural logarithm of the pure temperature effect $\Delta E_T$ (see Fig. 14) measured for the indirect gap of $^{28}$Si vs. the log of the temperature. The points are shown to follow rather accurately the $T^4$ law [29].

Fig. 16 Dependence of the n=1 yellow exciton energy on temperature in the $1<T<30$ K low temperature range. The points are experimental [73]. The curve is a fit to the experimental data with the expression $AT^4$ for $T \leq 20$ K. Note that above 20 K deviations from the $T^4$ law appear.

Fig. 17 The points represent experimental data [74] for the shift of the yellow exciton gap of $Cu_2O$ with temperature. The solid line represents a fit to the experimental points with Eq. (21) using only $\xi$ as adjustable parameter.

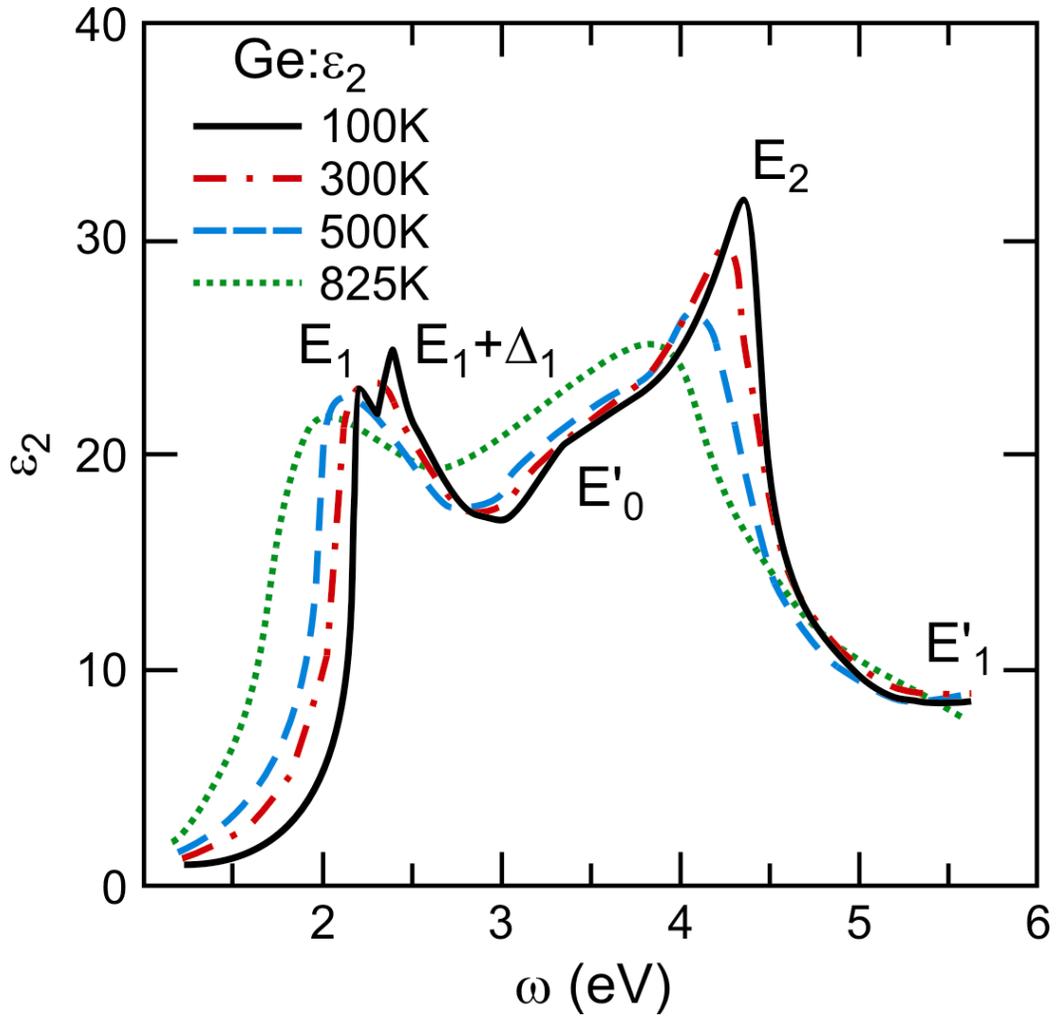

Figue 1

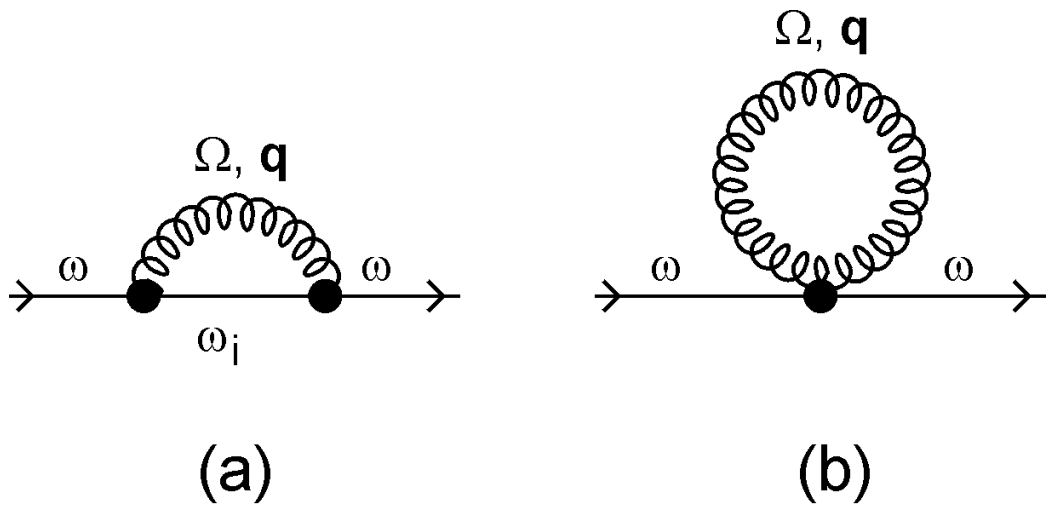

Figure 2

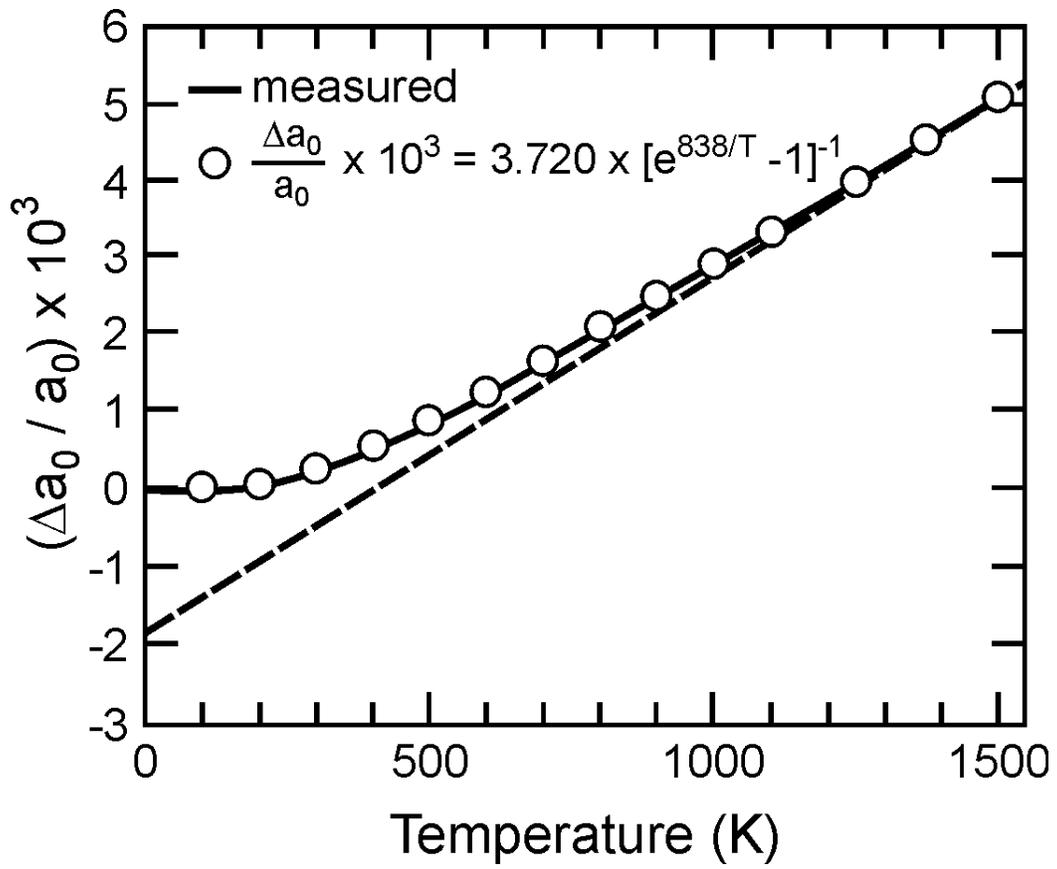

Figure 3
25

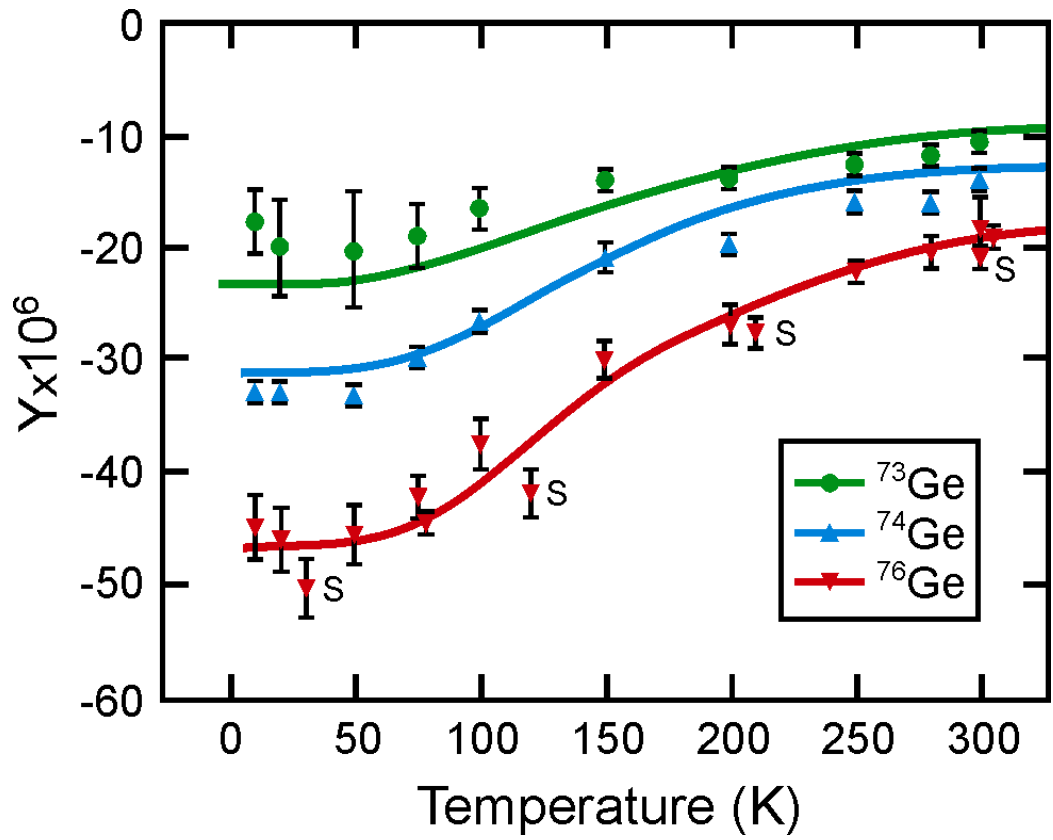

Figure 4

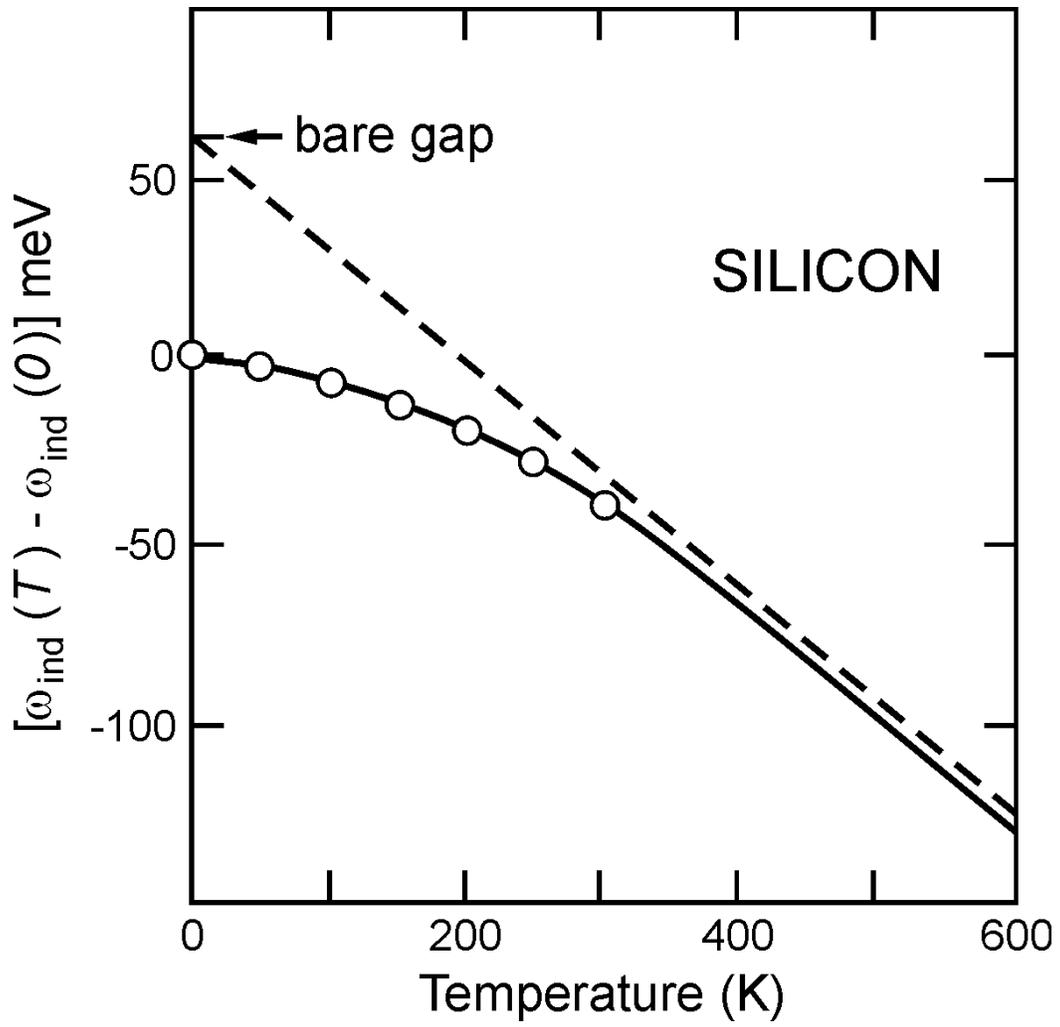

Figure 5

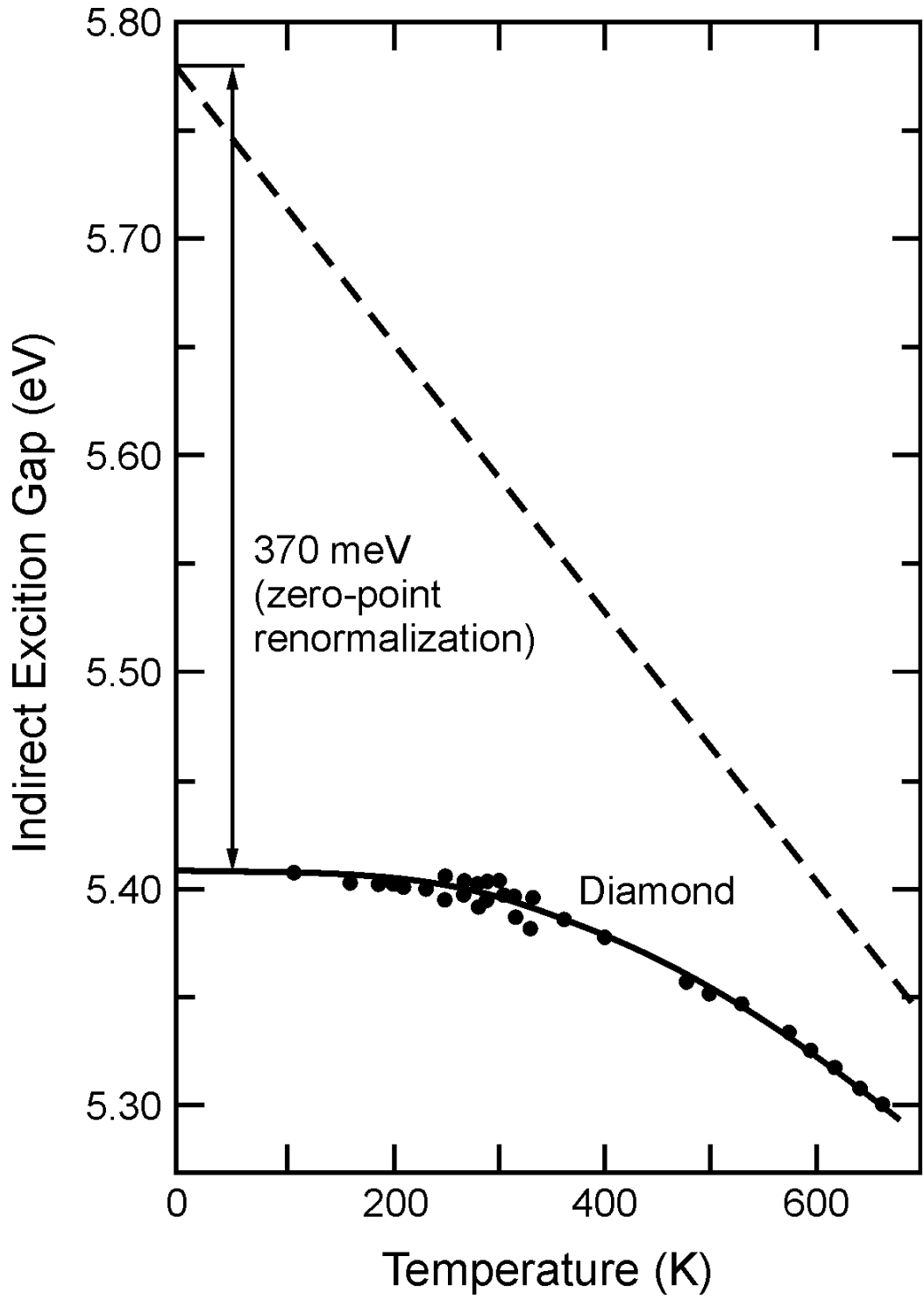

Figure 6

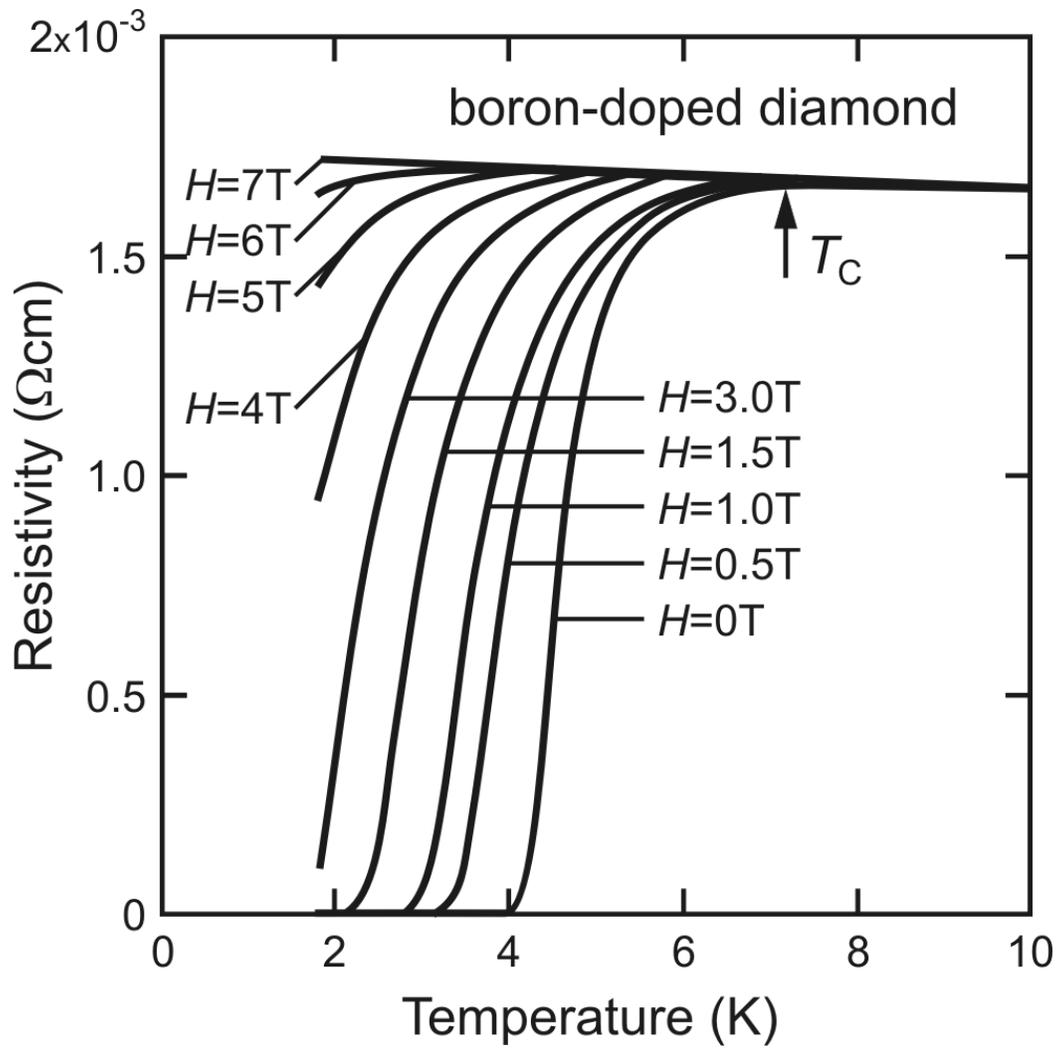

Figure 7

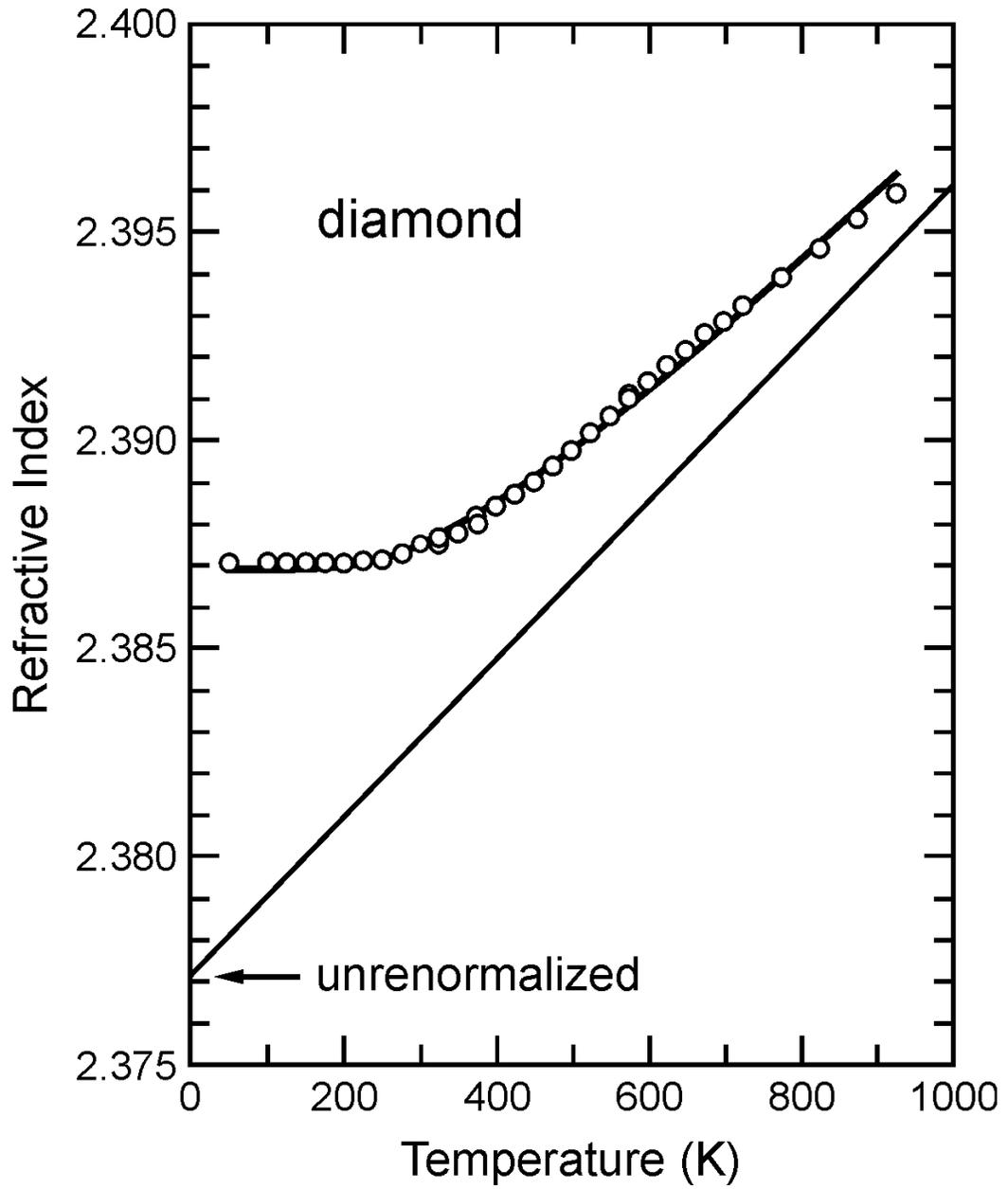

Figure 8

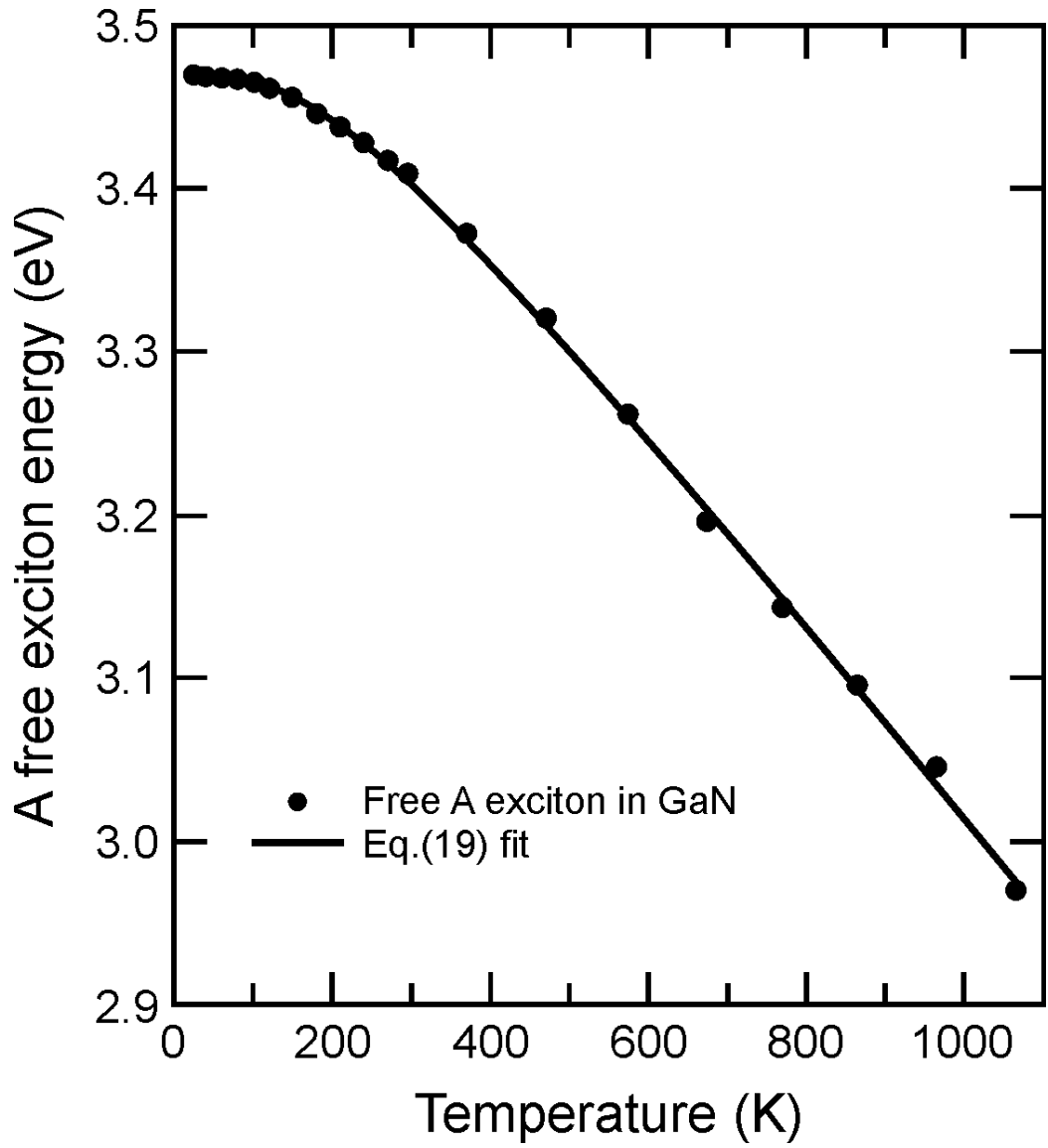

Figure 9

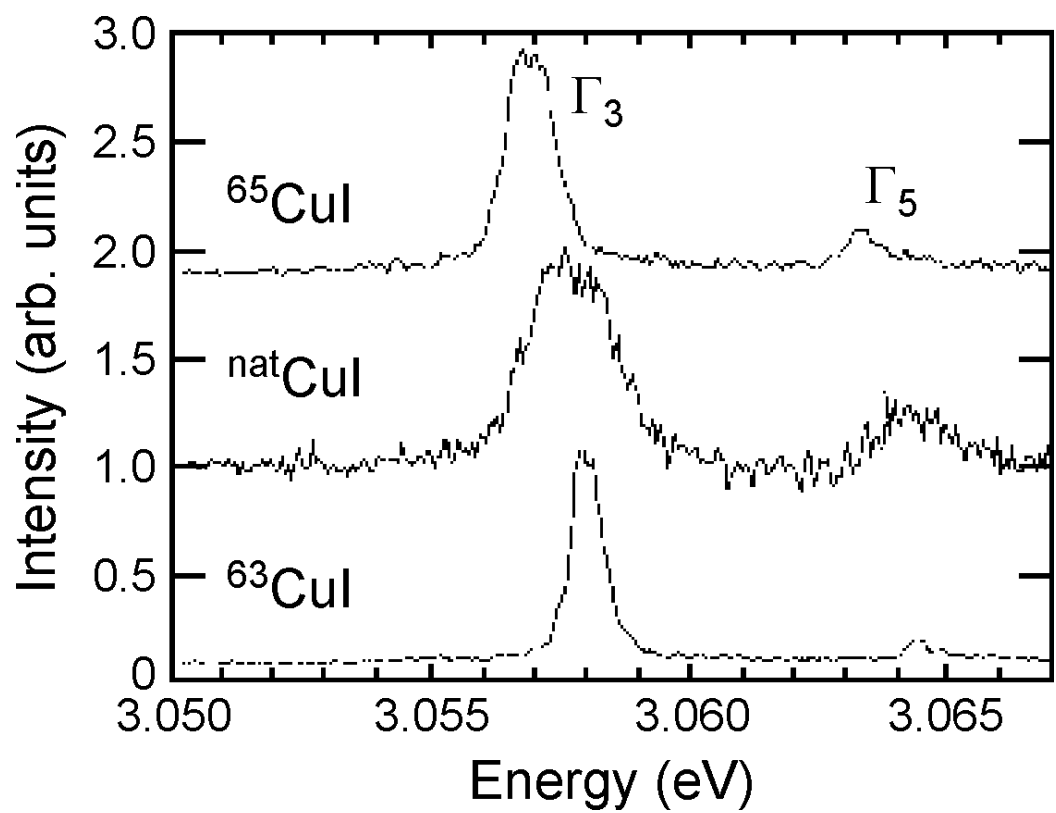

Figure 10

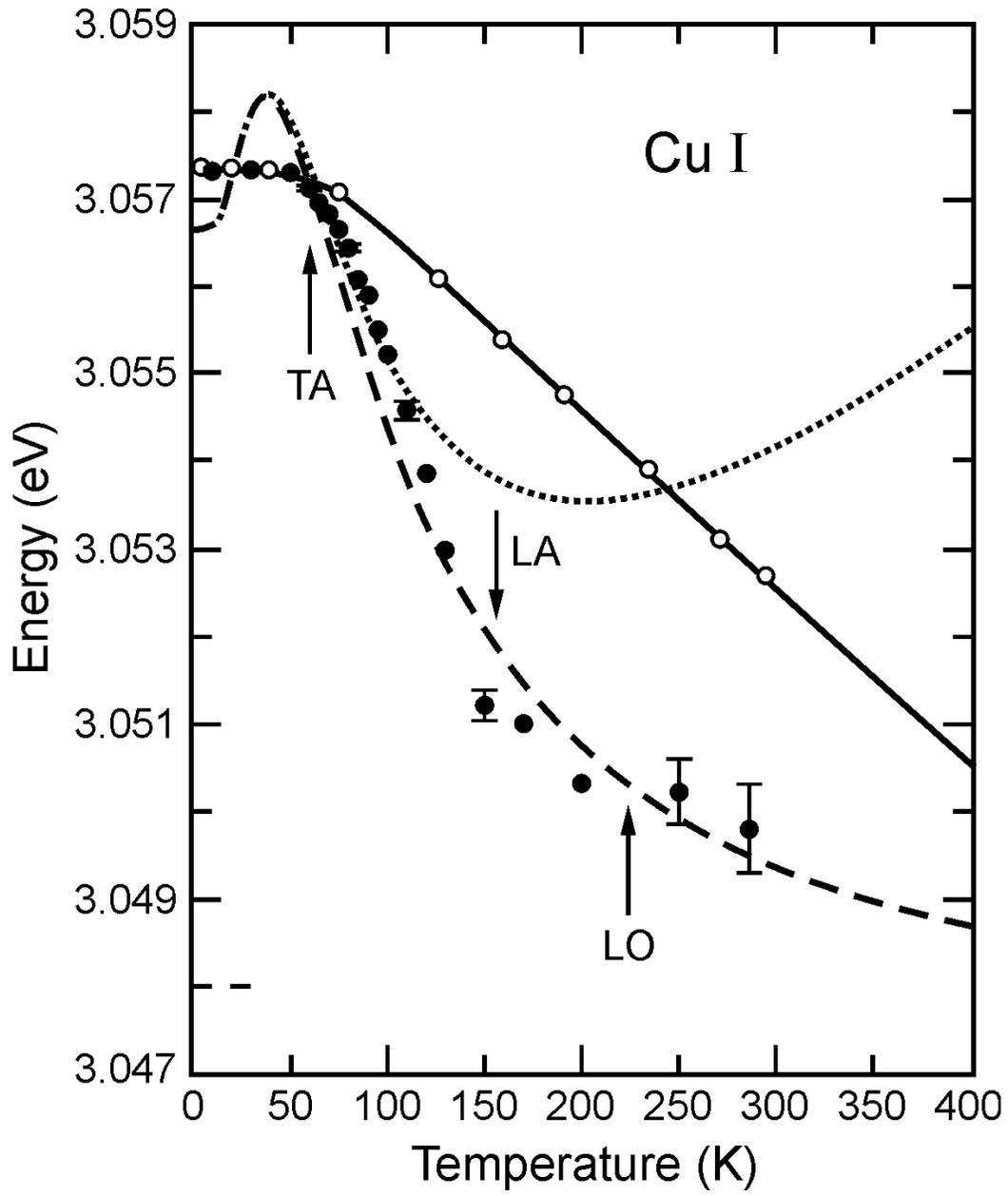

Figure 11

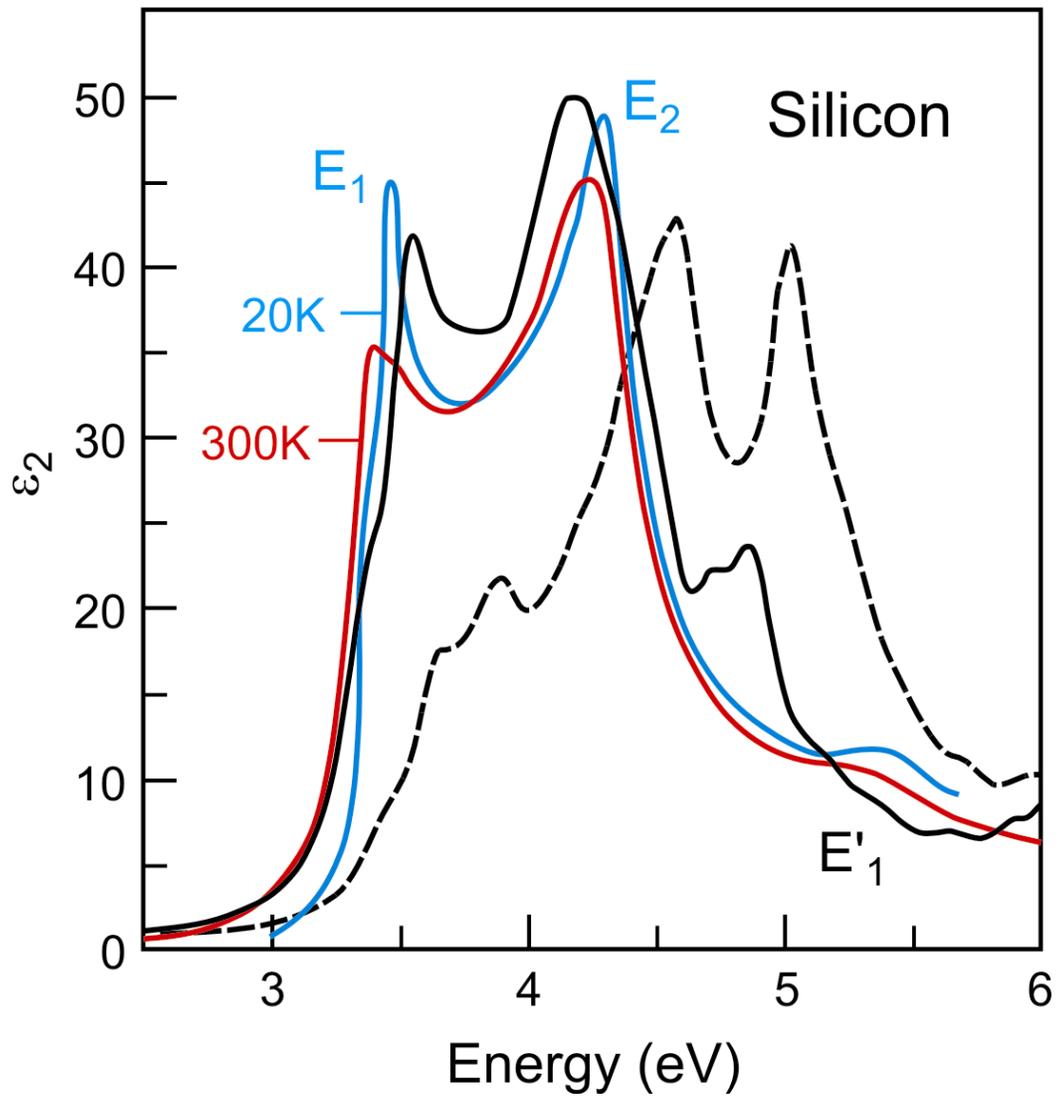

Figure 12

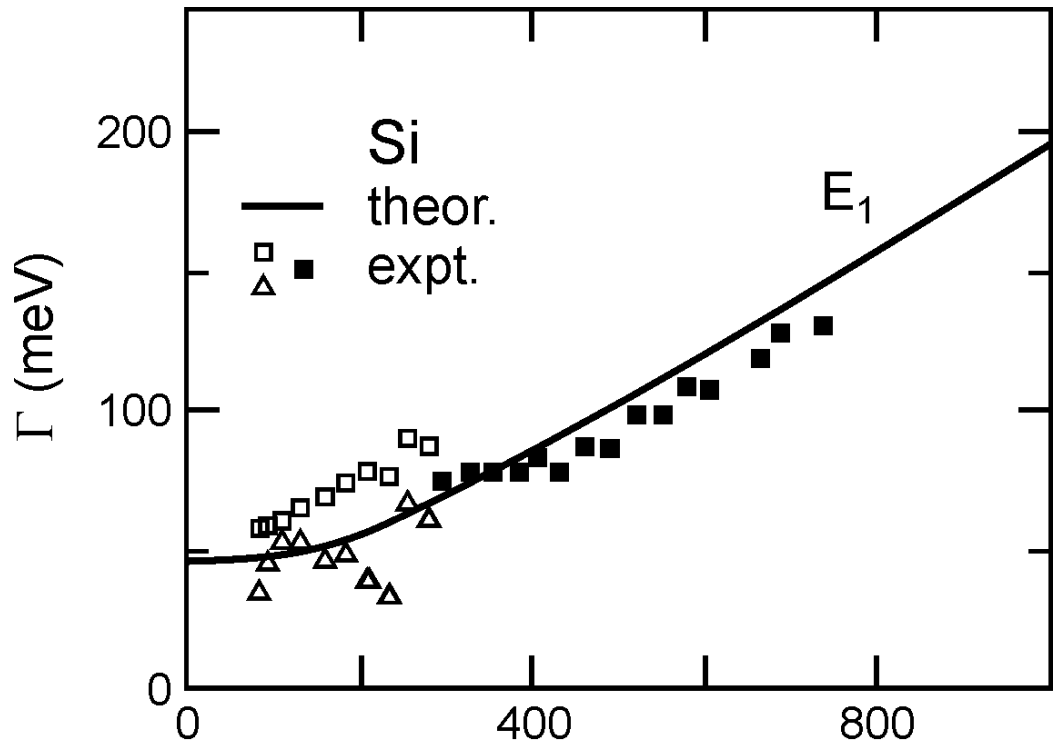

Figure 13

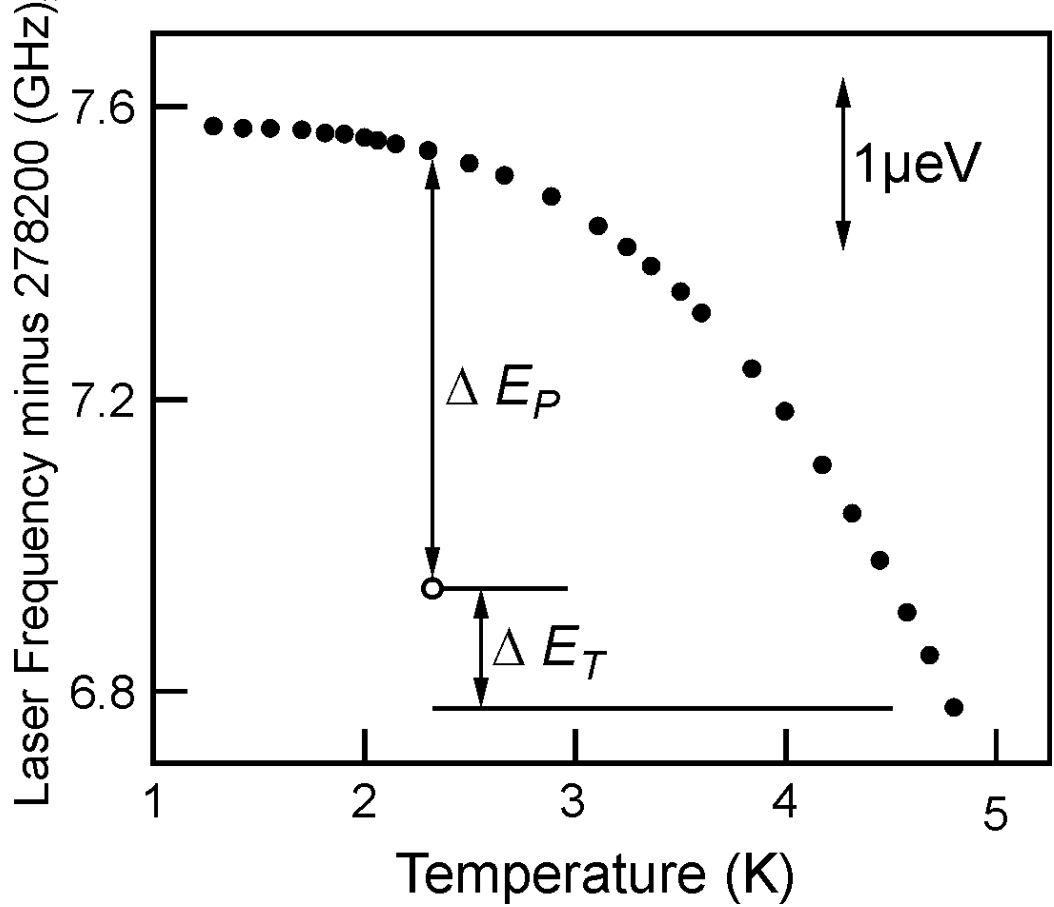

Figure 14

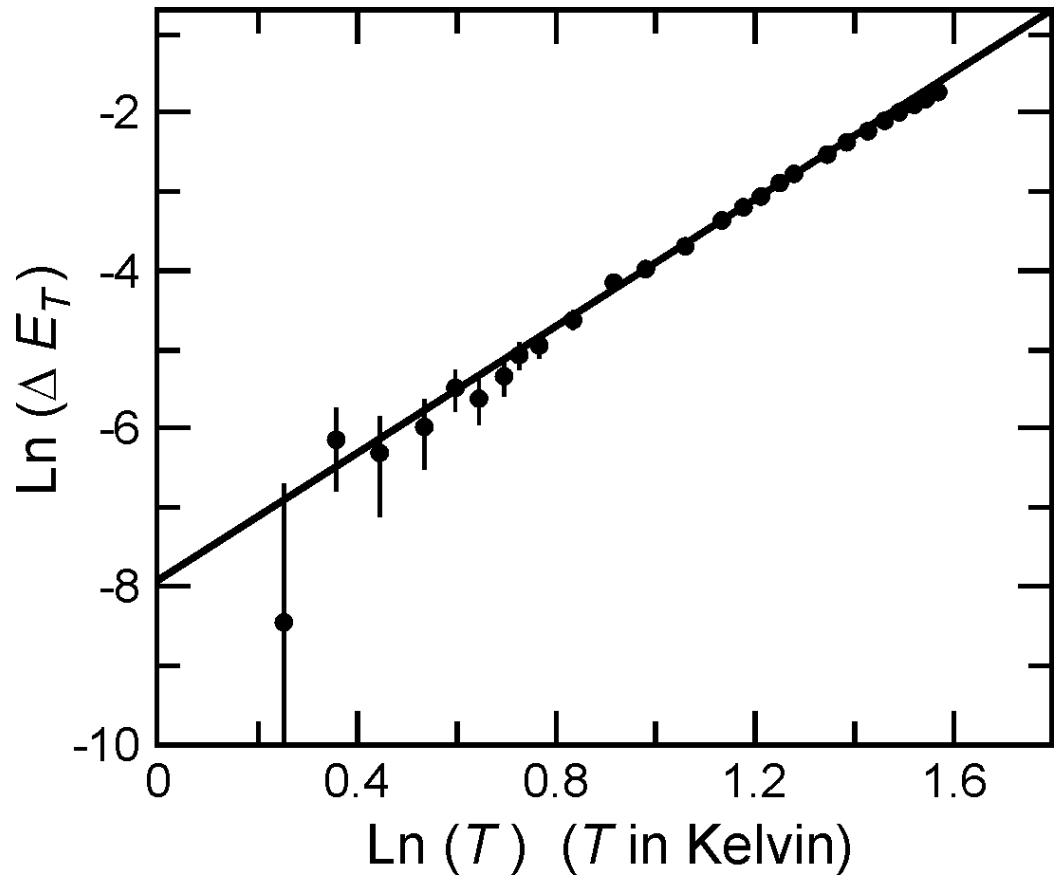

Figure 15

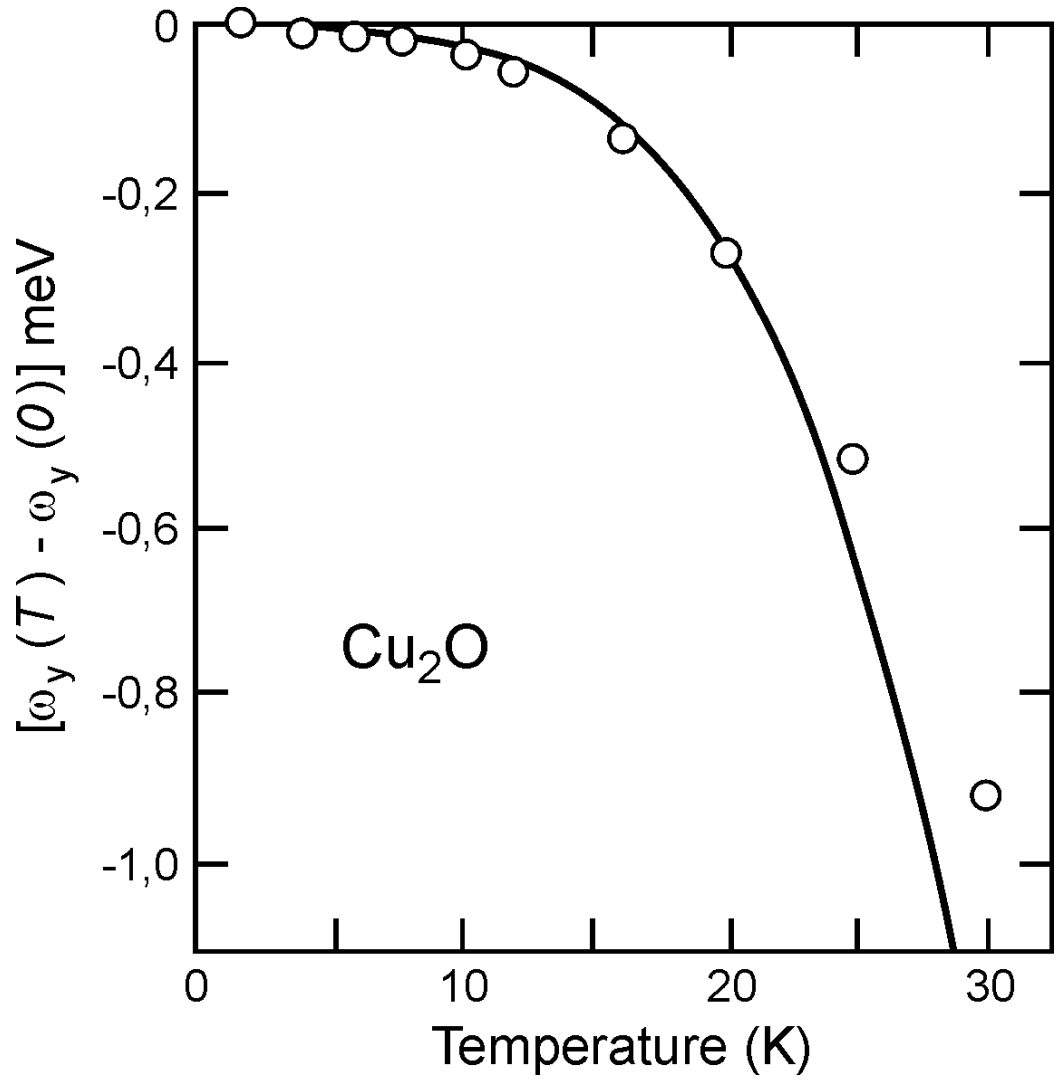

Figure 16

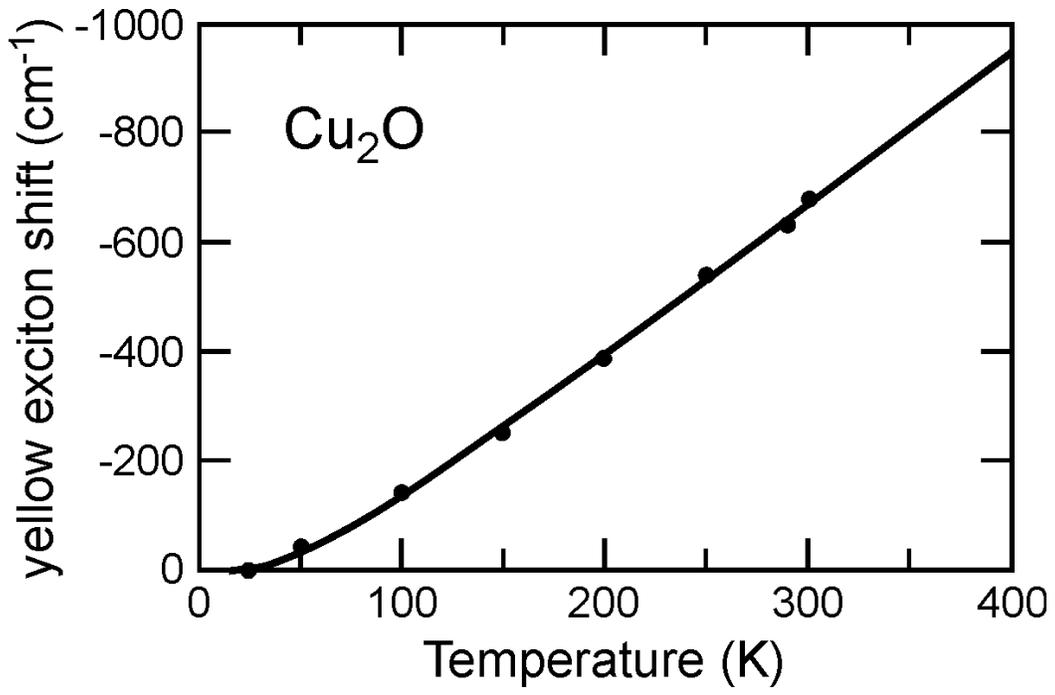

Figure 17